\documentclass[fleqn,usenatbib]{mnras}

\usepackage{newtxtext,newtxmath}
\usepackage[T1]{fontenc}

\DeclareRobustCommand{\VAN}[3]{#2}
\let\VANthebibliography\thebibliography
\def\thebibliography{\DeclareRobustCommand{\VAN}[3]{##3}\VANthebibliography}

\usepackage{graphicx}	% Including figure files
\usepackage[dvipsnames]{xcolor}
\usepackage{newtxtext,newtxmath}
\newcommand{\uv}{\textit{uv}}

\title[Inner walls or vortices?]{Inner walls or vortices? Crescent-shaped asymmetries in ALMA observations of protoplanetary discs}

\author[Á. Ribas et al.]{
Álvaro Ribas,$^{1}$\thanks{E-mail: ar2193@cam.ac.uk}
Cathie J. Clarke,$^{1}$
and Francesco Zagaria$^{1}$
\\
$^{1}$Institute of Astronomy, University of Cambridge, Madingley Road, Cambridge, CB3 0HA, UK 
}

\date{Accepted 2024 June 17. Received 2024 June 10; in original form 2024 March 20}

\pubyear{2024}

\begin{document}
\label{firstpage}
\pagerange{\pageref{firstpage}--\pageref{lastpage}}
\maketitle

\begin{abstract}
Crescent-shaped asymmetries are common in millimetre observations of protoplanetary discs and are usually attributed to vortices or dust overdensities. However, they often appear on a single side of the major axis and roughly symmetric about the minor axis, suggesting a geometric origin. In this work, we interpret such asymmetries as emission from the exposed inner cavity walls of inclined discs and use them to characterise their vertical extent. Here we focus on the discs around CIDA~9 and RY~Tau, first modelling their observations in visibility space with a simple geometric prescription for the walls, and then exploring more detailed radiative transfer models. Accounting for the wall emission yields significantly better residuals than purely axisymmetric models, and we estimate the dust scale height of these systems to be 0.4\,au at 37\,au for CIDA~9 and 0.2\,au at 12\,au for RY~Tau. Finally, we identify crescent-shaped asymmetries in twelve discs, nine of which have constraints on their orientation – in all cases, the asymmetry appears on the far-side of the disc, lending support to the hypothesis that they are due to their inner rims. Modelling this effect in larger samples of discs will help to build a statistical view of their vertical structure.
\end{abstract}


%%%%%%%%%%%%%%%%%%%%%%%%%%%%%%%%%%%%%%%%%%%%%%%%%%

%%%%%%%%%%%%%%%%% BODY OF PAPER %%%%%%%%%%%%%%%%%%

\section{Introduction}

In the last decade, observational studies of protoplanetary discs have revealed a plethora of substructures in them \citep[e.g.,][]{HLTau, Andrews2018_DSHARP}: rings, gaps, and azimuthal asymmetries (e.g., spiral arms and vortices) are ubiquitous in discs, and are likely to play a major role in planet formation through the concentration and retention of dust grains \citep[e.g.,][]{Pinilla2012, Carrera2021}. Thanks to observatories and instruments such as ALMA, VLA, SPHERE, or GPI, the radial structures of dozens of discs are now well characterised down to spatial scales of a few au, in many cases at multiple wavelengths \citep[e.g.,][]{Macias2021,Sierra2021,Bae2023,Benisty2023}.

In addition to the obvious importance of radial structures in planet formation, our understanding of this process also depends critically on the vertical distribution of solids in discs: if the observed substructures are caused by planets, then these must form quickly to explain the gaps and rings already present in discs at 1-2\,Myr. Mechanisms such as the streaming instability and pebble accretion are usually invoked to explain such fast formation \citep[e.g.,][]{Youdin2005,Johansen2017}, which rely on the settling and concentration of dust in the midplane. However, despite its key role and in contrast with their radial configuration, the vertical aspect of discs remains less understood for a number of reasons: protoplanetary discs are geometrically thin, with typical aspect ratios of $0.1-0.2$ for the gas and small grains in the disc atmosphere \citep[e.g.,][]{Avenhaus2018,Law2022} and much flatter distributions for large grains in the midplane \citep{Villenave2020, Villenave2022}, as expected from dust settling \citep[e.g.,][]{Dubrulle1995,Fromang2009}. This implies that the scales relevant for the vertical structure of discs can be quite small ($\lesssim{\rm au}$). Observing discs with moderate/high inclinations (a condition required to probe their vertical features) induces projection effects, combining emission at different radii and heights along the line of sight. Likewise, uncertainties in the optical depth and dust properties also hinder our ability to correctly interpret these data in terms of vertical structures.

Despite the aforementioned difficulties, a number of methods can be used to probe the vertical distribution of the various disc components. Scattered-light observations in the optical and near-infrared (IR) trace stellar photons scattered by $\sim\micron$-sized grains in the disc atmosphere, revealing the disc surface and enabling the measurement of its flaring \citep[e.g.,][]{Avenhaus2018}. In the same wavelength range, interferometric studies have reconstructed the emission from the very inner regions of discs, sampling the radial and vertical structure of their inner rim \citep{Lazareff2017,Kluska2020}. High-angular resolution observations with ALMA can probe the emitting surface of multiple gas molecules (even disentangling the upper and lower disc surfaces in some cases), which can then be used to determine the local conditions at different heights above the midplane \citep[][]{Pinte2018,Izquierdo2021,Law2023}. The contrast and extent of rings observed at millimetre wavelengths (tracing mm/cm-sized grains) can also be compared with models to constrain the level of turbulence and settling in discs \citep{Pinte2016,Pizzati2023}. Additionally, the unique orientation of edge-on discs offers a privileged look at their vertical configuration and the spatial segregation of different dust grain sizes \citep[e.g.,][]{Duchene2010, Wolff2017,Villenave2020,Villenave2022,Duchene2024,Villenave2024}. Combined, these studies are building a comprehensive understanding of the vertical structure of protoplanetary discs, but significant effort is still needed to match its radial counterpart.

\begin{figure}
	\includegraphics[width=\hsize]{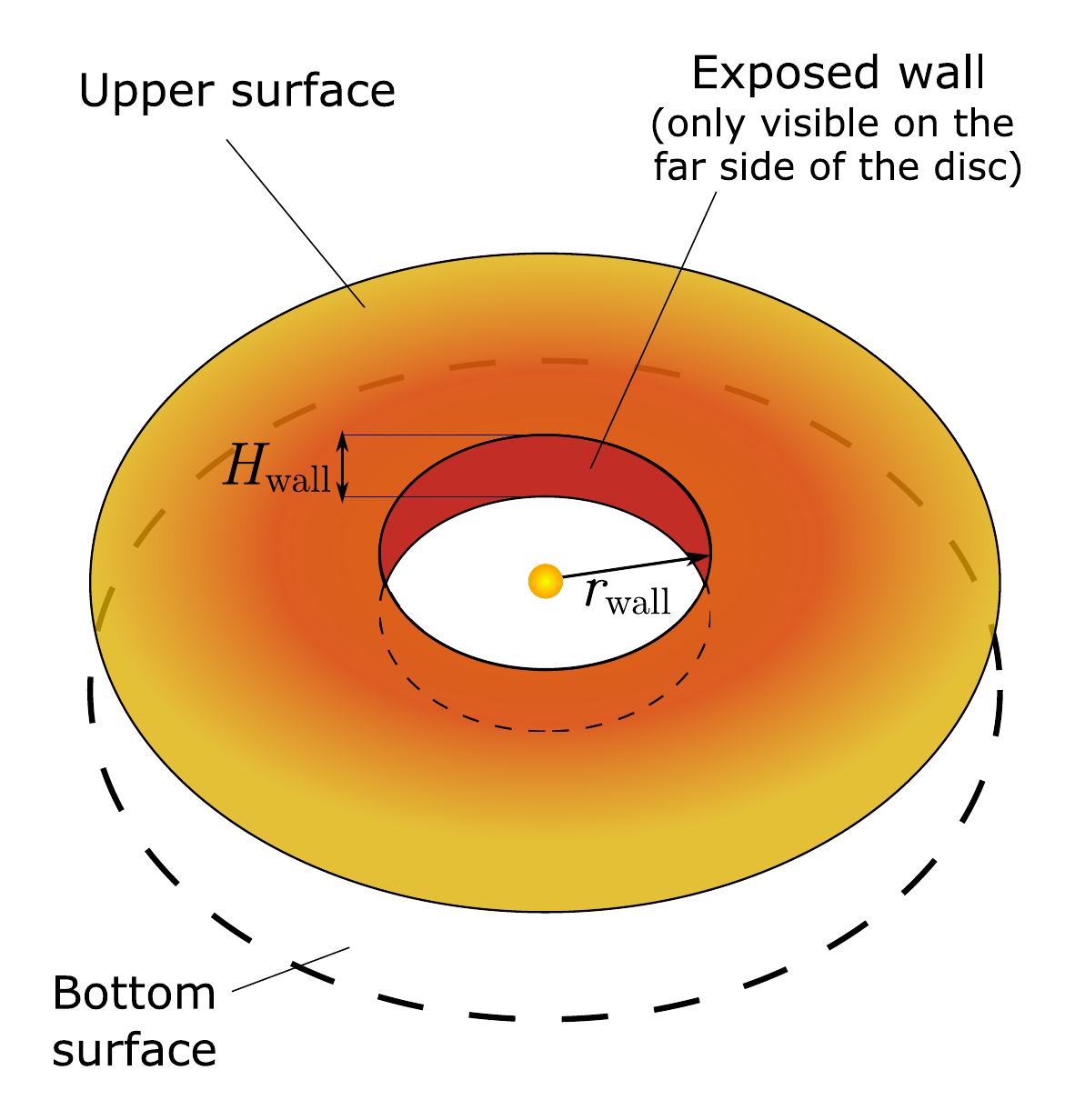}
    \caption{Schematic representation of the adopted geometric model. The hot inner wall is visible on the far side of the disc only (for an optically thick disc), creating an asymmetry along the minor axis. The modelling in Sec.~\ref{sec:modelling} aims at constraining $r_{\rm wall}$, $H_{\rm wall}$, and the flux $F_{\rm wall}$ arising from the wall.}
    \label{fig:scheme}
\end{figure}

Here, we investigate an alternative method to study the disc vertical extent through crescent-shaped asymmetries observed in discs cavities at (sub)millimetre wavelengths. In a protoplanetary disc, radiation from the central star reaches the disc inner wall with a more perpendicular angle than the disc surface and, as a result, the wall becomes hotter than the surrounding disc \citep[which has been proposed as the reason for the near-IR bump and possibly puffed-up inner discs of Herbig stars, e.g.][]{Natta2001, Dullemond2001, Dalessio2005, Dullemond2010}. This effect has been mostly explored in the optical/near-IR range, but it may also be detectable in the (sub)millimetre: \emph{if a disc is observed with some inclination and is optically thick at the cavity radius, its inner wall may appear as a crescent-shaped bright asymmetry on the far side of the disc only (where the wall is exposed, see Fig.~\ref{fig:scheme}), symmetrical about the minor axis on one side of the major axis}. In fact, crescent-shaped asymmetries are commonly observed in protoplanetary discs at these wavelengths (and preferentially at the edge of the cavity when observed with sufficient resolution), but they are typically interpreted as vortices \citep[e.g.,][]{vanderMarel2013,Isella2013,Cazzoletti2018,Boehler2021,Harsono2024} or overdensities due to eccentricity driven by massive planetary or binary companions \citep{Ragusa2017, Ragusa2020, Dong2018_MWC758}. However, a high number of them appear symmetric about the disc minor axis (see Sec.~\ref{sec:discussion}), which instead may suggest a geometric origin such as the one proposed here (at least in some cases).

In this work, we focus on ALMA observations of CIDA~9 and RY~Tau (two protoplanetary discs with crescent-shaped asymmetries) and model them to constrain their vertical structure. We begin by introducing the two targets and the processing of the ALMA data in Section~\ref{sec:sample_and_observations}. In Section~\ref{sec:modelling} we first describe our modelling of the observed visibilities using a simple geometric model for the disc wall, perform a comparison with more detailed radiative transfer models, and present the results for these to sources. Finally, the implications and interpretation of crescent-shaped asymmetries in ALMA observations are discussed in Section~\ref{sec:discussion}.

%%%%%%%%%%%%%%%%%%%%%%%%%%%%%%%%%%%%%%%%%%%%%%%%%%%%%%%%%%%%%
% SAMPLE AND OBSERVATIONS
\section{Sample and processing of ALMA data}\label{sec:sample_and_observations}

To test the possibility of constraining the vertical extent of inner rims at (sub)mm wavelengths, we first searched for discs showing clear crescent-shaped asymmetries aligned with their minor axis in ALMA observations. We then decided to focus on CIDA~9 and RY~Tau as two such examples of nearby and relatively bright discs which have been observed with different angular resolutions (by a factor of $\sim$3). While the asymmetry is clearly visible in both cases, this also allows us to explore the impact of resolution in our analysis.

\subsection{CIDA~9}

\begin{figure*}
    \centering
	\includegraphics[width=\columnwidth]{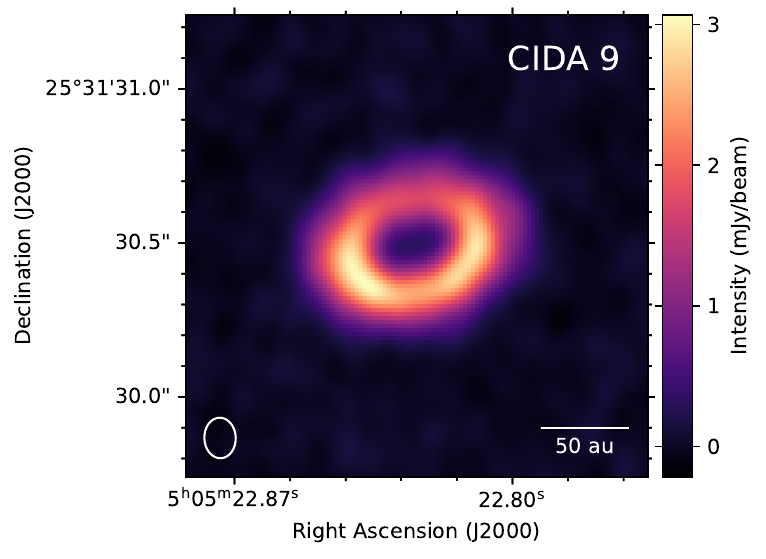}
    \includegraphics[width=\columnwidth]{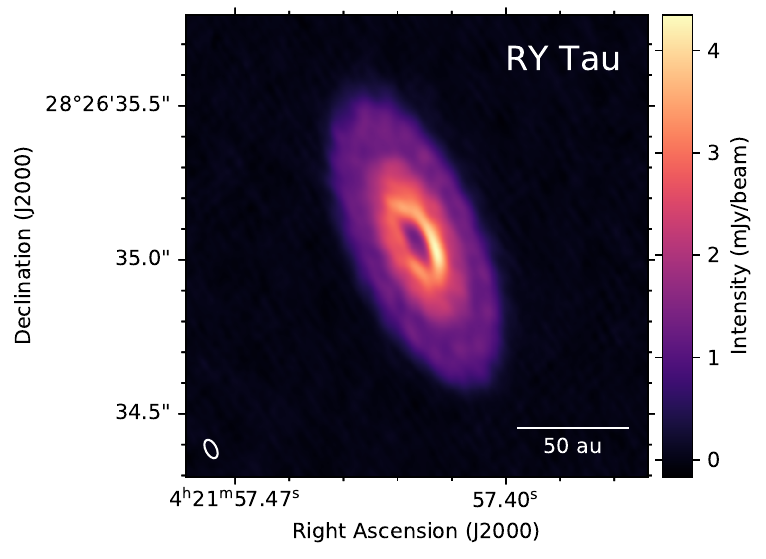}
    \caption{ALMA 1.3\,mm continuum images of the two sources studied in this work, CIDA~9 (left) and RY~Tau (right). The images were synthesised in CASA using {\tt tclean} with a {\tt robust} value of 0.5. The corresponding beams are shown as white ellipses in the bottom left corners. The southern side of CIDA~9 shows an increased emission with respect to the northern side. In the case of RY~Tau, the emission near the inner gap appears brighter on the West than on the East. This asymmetry appears aligned with the disc minor axis in both cases.}
    \label{fig:synthesized_images}
\end{figure*}

CIDA~9 (IRAS~05022+2527) is a wide binary system (separation $\sim$2.3\arcsec) located in the Taurus star-forming region \citep[$d$=175\,pc,][]{GaiaDR3}. The main component, CIDA~9A, is an M1.8 star \citep{Herczeg2014} surrounded by a disc with a $\sim$40\,au cavity, as probed by ALMA 1.3\,mm continuum observations \citep{Long2018}. These data already showed an azimuthal asymmetry, with the southern side of the disc being the brightest and the maximum intensity peak located on the South-East part of the disc (see Fig.~\ref{fig:synthesized_images}). More recently, \citet{Harsono2024} presented additional ALMA observations at 3\,mm which revealed a similar morphology at this longer wavelength. They also identified a tentative shift in the peak position between the 1.3\,mm and the 3\,mm observations and suggested that a vortex may explain the asymmetry. Here we focus on the 1.3\,mm observations as they have better sensitivity.

Based on previous studies, we adopted a stellar temperature of 3590\,K and a stellar radius of 1.2\,$R_\odot$ from \citet{Herczeg2014} (after updating with the new Gaia distance) for CIDA~9, as well as the stellar mass of 0.6\,$M_\odot$ derived in \citet{Harsono2024} by fitting the $^{12}$CO (2-1) Keplerian rotation of the disc.

\subsubsection{Processing of ALMA data for CIDA~9}

We downloaded the CIDA~9 observations at 1.3\,mm from the ALMA archive (project 2016.1.01164.S, P.I.: Herczeg). These observations were taken on 31 August 2017, with baselines ranging from 20\,m to 3.6\,km. The correlator setup included two spectral windows for continuum centred at 218.0 and 233.0\,GHz, with a bandwidth of 1.875 GHz each. The setup also contained three spectral windows covering CO isotopologues with narrower bandwidths, which we did not use in this study. We first calibrated the data using the pipeline scripts with CASA 5.1.1, and then applied three rounds of phase-only self-calibration with decreasing time intervals using CASA 6.5.1. This process increased the peak S/N by 20\,\%. The companion CIDA~9B (2.3\arcsec North-East of CIDA~9) is also detected in the observations 2.3\,arcsec North-East of CIDA~9, and it is $\sim$100 times fainter than CIDA~9A at 1.3\,mm. To avoid any potential impact in the modelling, we removed its contribution from the visibilities by subtracting a point source with the corresponding flux and position. Finally, we placed the phase centre at the measured location of the disc in the image (RA=05h05m22.82s, Dec=25d31m30.49s) and binned the visibilities using one channel per spectral window and 60s time bins. A synthesised image of CIDA~9 using a {\tt robust} value of 0.5 is shown in Fig.~\ref{fig:synthesized_images}, resulting in a 0.13\arcsec$\times$0.10\arcsec (PA=2$^{\circ}$) beam, a disc flux of 37$\pm$2\,mJy (including the absolute 5\,\% calibration uncertainty), a peak value of $3.4\,{\rm mJy\,beam^{-1}}$ and an rms of ${\rm 48\,\mu Jy\,beam^{-1}}$.

We note that there are additional observations of CIDA~9 at 1.3\,mm in the ALMA archive (project 2018.1.00771.S, P.I.: Manara) with a slightly lower angular resolution (0.16$\times$0.10\arcsec). We initially downloaded, processed, and combined them with the data described above. However, while this process resulted in an improved S/N, it also increased the number of visibility measurements, making the geometric modelling in Sec.~\ref{sec:modelling_process} significantly slower. Given that the S/N of the observations from project 2016.1.01164.S is already high and the data in 2018.1.00771.S do not improve the recoverability of small scales, we did not include them in our analysis.

\subsection{RY~Tau}

RY~Tau is an F7-G1 star \citep[e.g.,][]{Calvet2004,Herczeg2014} also located in the Taurus star-forming region (although distance estimates are complicated by its surrounding nebulosity). The intermediate resolution ($\sim$0.12\arcsec) 1.3\,mm ALMA data presented in \citet{Long2018} already revealed an inner cavity in the disc, as well as a crescent-shaped asymmetry on the North-West side. Later ALMA observations at higher angular resolution (0.04\arcsec) showed a clear azimuthal asymmetry, with an increase in brightness in the inner part of the disc along the North-West minor axis \citep[Fig.~\ref{fig:synthesized_images}, also see][]{Francis2020}. At shorter wavelengths, RY~Tau is known to have a jet \citep{Garufi2019}, displays signatures of a possible disc wind in scattered light \citep{Valegard2022}, and near-IR interferometric observations also suggest the presence of an inner disc in the system \citep{Davies2020}.

While a range of stellar parameters have been reported for RY~Tau, in this work we adopted those in \citet{Valegard2022} \citep[also compatible with][]{Davies2020}, i.e., a temperature of 5945\,K, a radius of 3.25\,$R_\odot$, and a mass of 1.95\,$M_\odot$. Regarding the distance, \citet{GaiaDR3} places RY~Tau at 138\,pc, but this estimate is significantly uncertain due to the nebulosity around it. We maintained this value for the distance, as it is in agreement with its location in the Taurus region and allows for comparisons with previous works which have historically adopted values of 130-140\,pc \citep[e.g.,][]{Garufi2019,Davies2020,Valegard2022}.

\subsubsection{Processing of ALMA data for RY~Tau}

We used two different data sets for RY~Tau to achieve both high angular resolution and sufficient coverage at short baselines. For the compact configuration, we use observations from project 2016.1.01164.S on 27 August 2017 (the same ALMA program than the CIDA~9 observations). The baseline range and spectral setup of the correlator were identical to those of CIDA~9. As explained in \citet{Long2018}, these observations required additional processing due to an issue with the spectral configuration of the phase calibrator, and we used the self-calibrated visibilities provided by these authors.

We also downloaded RY~Tau observations from project 2017.1.01460.S (P.I.: Hashimoto), which were taken on 08 October 2017 with baselines ranging from 40\,m to 16.1\,km. The correlator setup included two spectral windows for continuum (central frequencies of 215.0 and 232.6\,GHz, 1.875\,GHz bandwidth) as well as other three targeting CO isotopologues with significantly less bandwidth, which we did not use for our analysis. As in the case of CIDA~9, the data were first calibrated with CASA 5.1.1 and two rounds of phase self-calibration were applied, increasing the peak S/N by 50\,\%. We then performed a round of joint self-calibration of the compact and extended data sets to ensure proper alignment. The disc fluxes measured from each observation are compatible within 1\,\% and thus we did not rescale them. Finally, we shifted the phase centre to the approximate centre of the disc (RA=04h21m57.422s Dec=28d26m35.045s) and binned the visibilities using one channel per spectral window and 60s time bins. Figure~\ref{fig:synthesized_images} shows the synthesised image of RY~Tau using a {\tt robust} parameter of 0.5, with a 0.06\arcsec$\times$0.035\arcsec (PA=27$^{\circ}$) beam, a disc flux of 217$\pm$11\,mJy (including a 5\,\% absolute calibration uncertainty), a peak value of $3.3\,{\rm mJy\,beam^{-1}}$, and an rms of $33.5\,{\rm \mu Jy\,beam^{-1}}$.

%%%%%%%%%%%%%%%%%%%%%%%%%%%%%%%%%%%%%%%%%%%%%%%%%%%%%%%%%%%%%
% MODELLING
\section{Wall modelling}\label{sec:modelling}

To constrain the vertical structure of CIDA~9 and RY~Tau based on their asymmetries, we first build a simple geometric model containing an azimuthally-symmetric surface brightness and the contribution of a flat inner wall. We model the observations in visibility space and constrain the vertical thickness of their inner rims using Bayesian analysis and a Monte Carlo Markov Chain (MCMC). We then turn to more sophisticated (and computationally expensive) radiative transfer models for additional information on the dust scale heights.

\subsection{Geometric modelling}\label{sec:geometric_model}

\subsubsection{Model setup}\label{sec:model_description}

We decided to test a simple geometric model for the wall on the observed visibilities instead of on synthesised images for two main reasons: firstly, the scales of the vertical structure of discs are small ($\lesssim$ au), and their analysis benefits from the highest angular resolution possible. While image synthesis greatly helps in interpreting interferometric observations, the direct analysis of visibilities provides the most information about small spatial scales, and super-resolution techniques using visibilities can identify structures 3-5 times smaller than the synthesised beam \citep[e.g.,][]{frank}. Moreover, visibilities have well-defined Normal uncertainties, in contrast with the correlated noise of reconstructed images.

\begin{figure*}
    \centering
	\includegraphics[width=0.8\hsize]{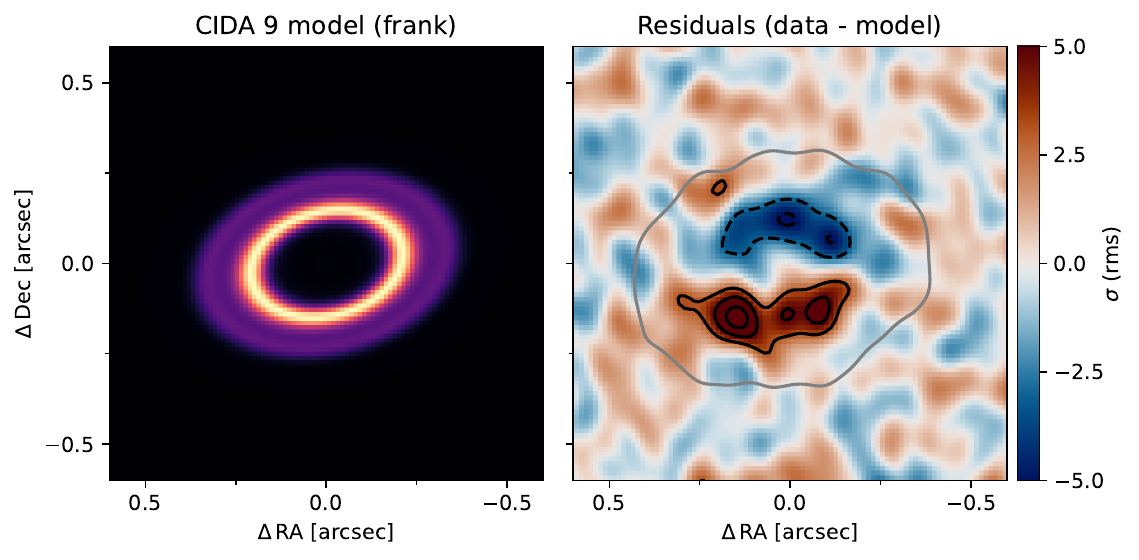}\
    \includegraphics[width=0.8\hsize]{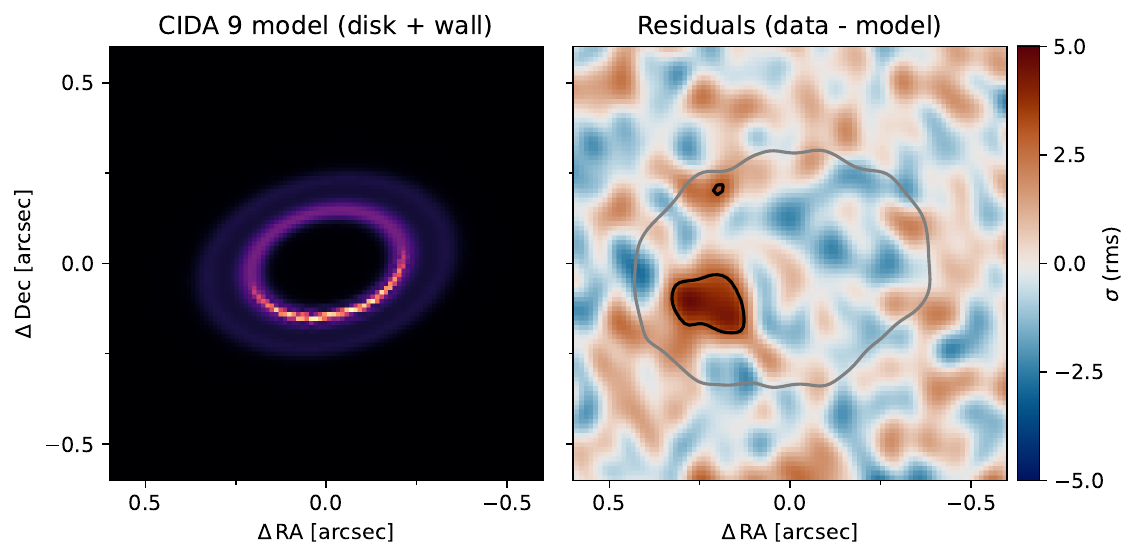}
    \caption{Comparison of an axisymmetric model and the geometric wall model for CIDA 9. Top row: axisymmetric model from {\tt frank} (left) and the corresponding residuals (right). Bottom row: geometric model including the wall emission (left) and the corresponding residuals (right). The colour scale of the models is normalised from the minimum to the maximum value in each case. The residuals were produced by imaging the residual visibilities, i.e., subtracting the model visibilities to the observed data and synthesising an image with a {\tt robust} value of 0.5. The residuals are shown in units of the image rms. The black contours represent 3, 5, and 7 times the rms (solid lines), and -3, -5, and -7 (dashed lines). The grey contour shows the 5 rms level of the observations.}
    \label{fig:residuals_CIDA9}
\end{figure*}

\begin{figure*}
    \centering
	\includegraphics[width=0.8\hsize]{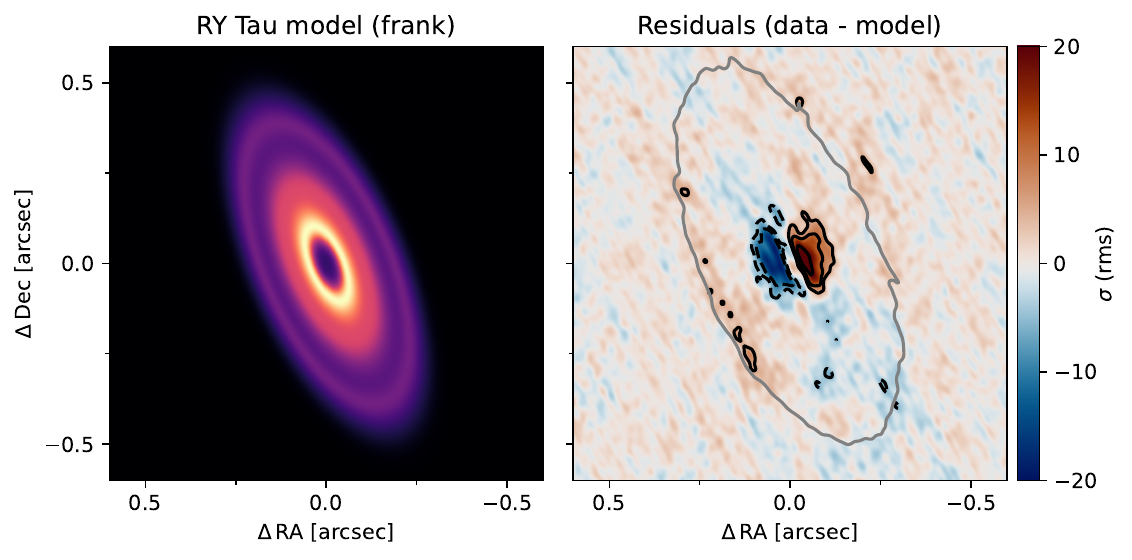}\
    \includegraphics[width=0.8\hsize]{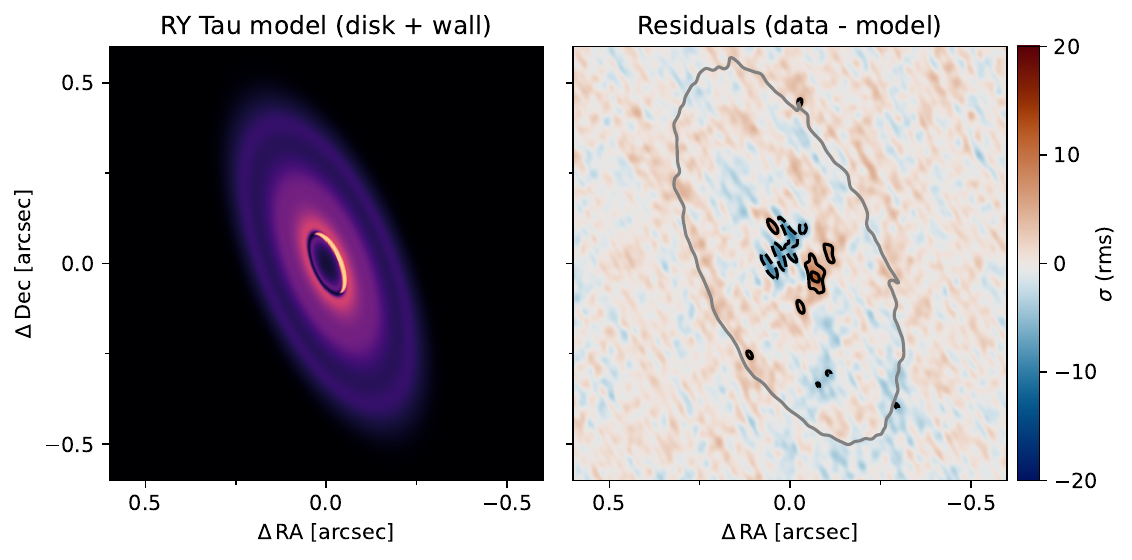}
    \caption{Comparison of the best fit axisymmetric disc model and the geometric model including emission from the inner wall for RY Tau. Figures are as in Fig.~\ref{fig:residuals_CIDA9}. In this case, the black contours correspond to 5, 10, and 20 times the image rms (solid lines), and -5, -10, and -20 (dashed lines).}
    \label{fig:residuals_RY_Tau}
\end{figure*}

We employed the {\tt frank} \citep{frank} and {\tt galario} \citep{galario} packages to model the observations. On the one hand, {\tt frank} assumes discs to be azimuthally symmetric and uses Hankel transforms to derive their intensity radial profile directly from the visibilities (i.e., non-parametrically). This method by-passes the need to adopt a specific functional form for the radial profile. In contrast, {\tt galario} computes the visibilities of any sky brightness distribution for a given set of \uv-coordinates, thus requiring an input model but relaxing the azimuthal symmetry condition. In our approach, we used {\tt galario} to calculate the visibilites of a 2D image containing an axisymmetric disc (produced with {\tt frank}) plus a wall, resulting in a non-axisymmetric brightness distribution. Our focus is in evaluating the wall properties and we are not concerned with the overall disc structure, so the radial profile from {\tt frank} is a convenient way to account for the disc emission without modelling it – this is particularly useful for discs with substructures that would otherwise require multiple free parameters to properly describe their radial profiles.

To calculate the disc plus wall model (hereafter the \emph{geometric model}), we first need to determine a pixel scale and image size appropriate for {\tt galario}. As described in \citet{galario}, the \uv-coverage of the observations and the antenna size impose constraints on these quantities to ensure both proper sampling of small scales and enough coverage of the field of view (FOV) to avoid aliasing of large-scale structures. Using {\tt galario}'s functionalities, we determined that CIDA~9 required a pixel size of 0.012\arcsec (2\,au at the source distance) and 2048 pixels. For RY~Tau we adopted a pixel scale of 0.0031\arcsec (0.4\,au) and 4096 pixels (this number of pixels is smaller than the FOV suggested by {\tt galario} by a factor of two, but it is $>$12 times larger than the disc diameter and allows for faster modelling).

We used {\tt frank} to obtain initial radial profiles of both discs from their observed visibilities. For {\tt frank}'s hyper-parameters, we adopted an $\alpha$ value of 1.05, $w_{\rm smooth}$ of 10$^{-4}$ and 10$^{-2}$ for CIDA~9 and RY~Tau, respectively, and performed the fit using a log-normal scale. We also tested other reasonable values for these parameters and restricted the fit to \uv-distances to $<$2\,M$\lambda$ for CIDA~9 and $<$6\,M$\lambda$ for RY~Tau to limit the impact of the wall asymmetry (traced by the most extended baselines), with no significant effect on the profiles. The discs' geometry (inclination $i$ and position angle ${\rm PA}$) were initially set to those reported in \citet{Long2018} and left to vary during the fitting. Since both sources contain azimuthal asymmetries, the resulting profiles are an incomplete description of the systems: at radii where wall emission is present, the radial profiles will predict intensities in between the disc and the wall true ones. This is visible in the imaged residuals after subtracting the visibilities of the {\tt frank} profiles from the observations \citep[see Figs.~\ref{fig:residuals_CIDA9} and~\ref{fig:residuals_RY_Tau}, also see the residuals in][]{Long2018, Jennings2022}. Therefore, the derived {\tt frank} profiles are only a first approximation, and we correct them later in the process by subtracting the wall emission at the corresponding radii.

The next step is to calculate the emission of the inner wall for a given set of disc and geometric parameters. To do this, we assumed a cavity radius of $r_{\rm wall}$ with a sharp wall of vertical thickness $H_{\rm wall}$\footnote{In this work, we use $H$ to refer to the thickness of the emitting inner rim at (sub)mm wavelengths, and $h$ for the pressure scale height.} (see Fig.~\ref{fig:scheme}). The disc is observed with an inclination $i$, and a total flux $F_{\rm wall}$ is received from the wall. To determine which pixels contain emission from the wall and deal with its edges (i.e., pixels only partially covering the wall), we built a finer grid with a pixel size of 0.05\,au and enough FOV to fully cover the cavity of each disc. Using this high-resolution grid, we calculated the ellipses traced by the cavity at the upper and lower disc surfaces: the visible rim corresponds to pixels belonging to the upper surface ellipse only (the area labeled "Exposed wall" in Fig.~\ref{fig:scheme}). The wall flux $F_{\rm wall}$ is evenly distributed across the masked pixels to determine their intensities. We then degrade this fine grid back to {\tt galario}'s resolution, resulting in a 2D image of the wall intensity $I_{\rm wall}$, as well as the fractional wall coverage ($f_{\rm wall}$) for each pixel in {\tt galario}'s grid.

The final step is to combine the emission from the wall and the disc, which requires two additional considerations. Firstly, as formerly mentioned, the initial {\tt frank} profiles need to be corrected at radii including wall emission. For this purpose, we subtracted a Gaussian ring centred at $r_{\rm wall}$, with a a total flux of $F_{\rm ring}$ and a width of $\sigma_{\rm ring}$. The corrected radial profile is projected onto the {\tt galario} grid with an inclination $i$. Finally, we assume the wall emission to be optically thick at these wavelengths, and thus the total intensity in each pixel is $I_{\rm wall} + I_{\rm disc} (1 - f_{\rm wall})$, where $I_{\rm wall}$ is the intensity of the wall, $I_{\rm disc}$ is the disc intensity, and $f_{\rm wall}$ is the wall coverage in that pixel.

\subsubsection{Fitting process}\label{sec:modelling_process}

We fit the observed visibilities with a Bayesian approach using the {\tt emcee} software \citep{emcee}. The model described in Sec.~\ref{sec:model_description} consists of six free parameters, namely $r_{\rm wall}$, $H_{\rm wall}$, $F_{\rm wall}$, $F_{\rm ring}$, $\sigma_{\rm ring}$, and $i$. To fit the observations, we also included the disc position angle ${\rm PA}$ and possible offsets from the image centre $\Delta \alpha$ and $\Delta \delta$. A global flux scaling factor $A$ was also introduced as a nuisance parameter to better match the total flux of the observations. Therefore, the full model contains 10 free parameters. For any given combination of free parameters $\theta$, we calculate the model image as described in Sec.~\ref{sec:model_description} and determine the visibilities at the observed \uv-coordinates using {\tt galario} to compare against the observations.

\begin{figure*}
    \centering
	\includegraphics[width=\columnwidth]{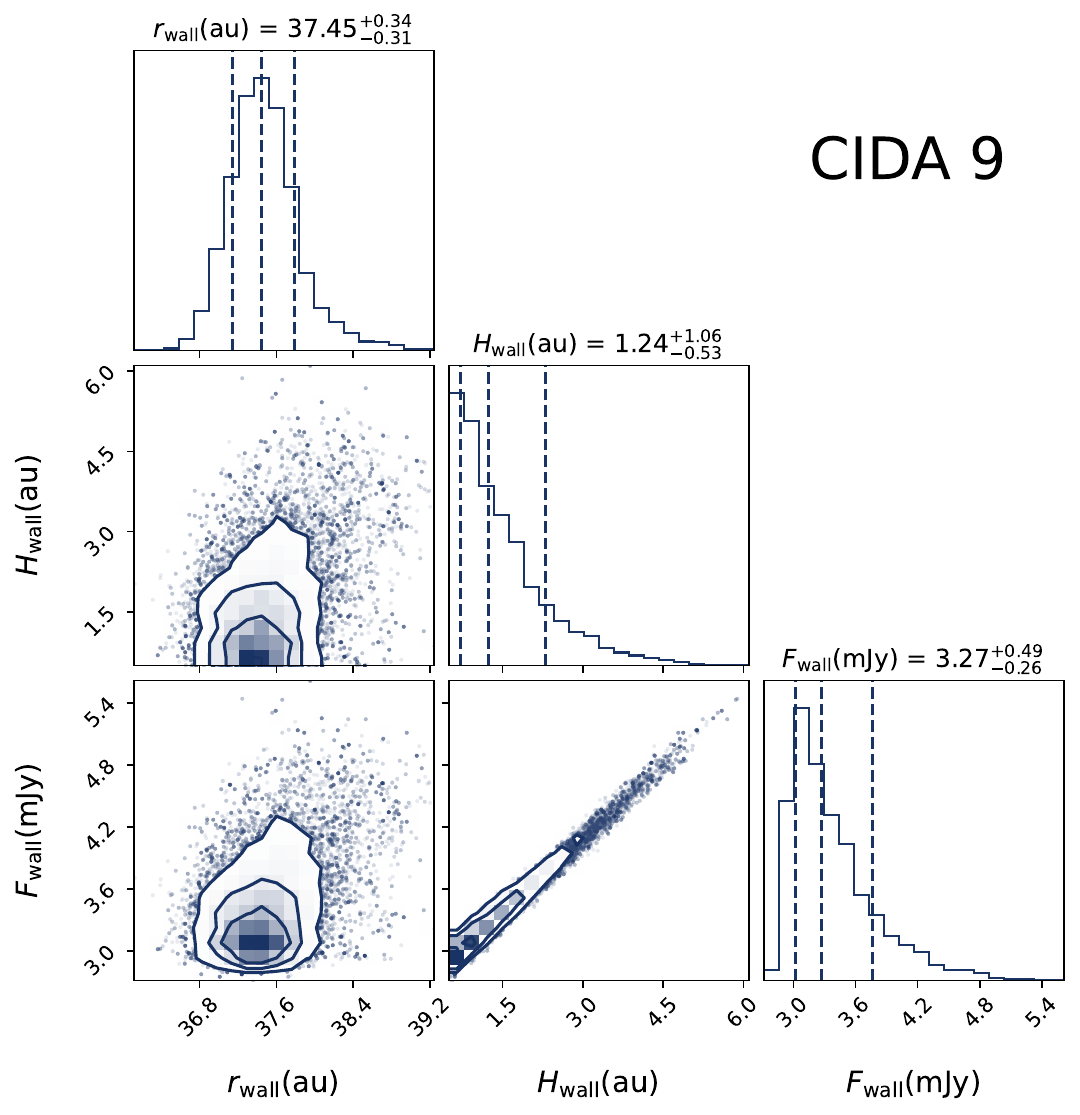}\
    \includegraphics[width=\columnwidth]{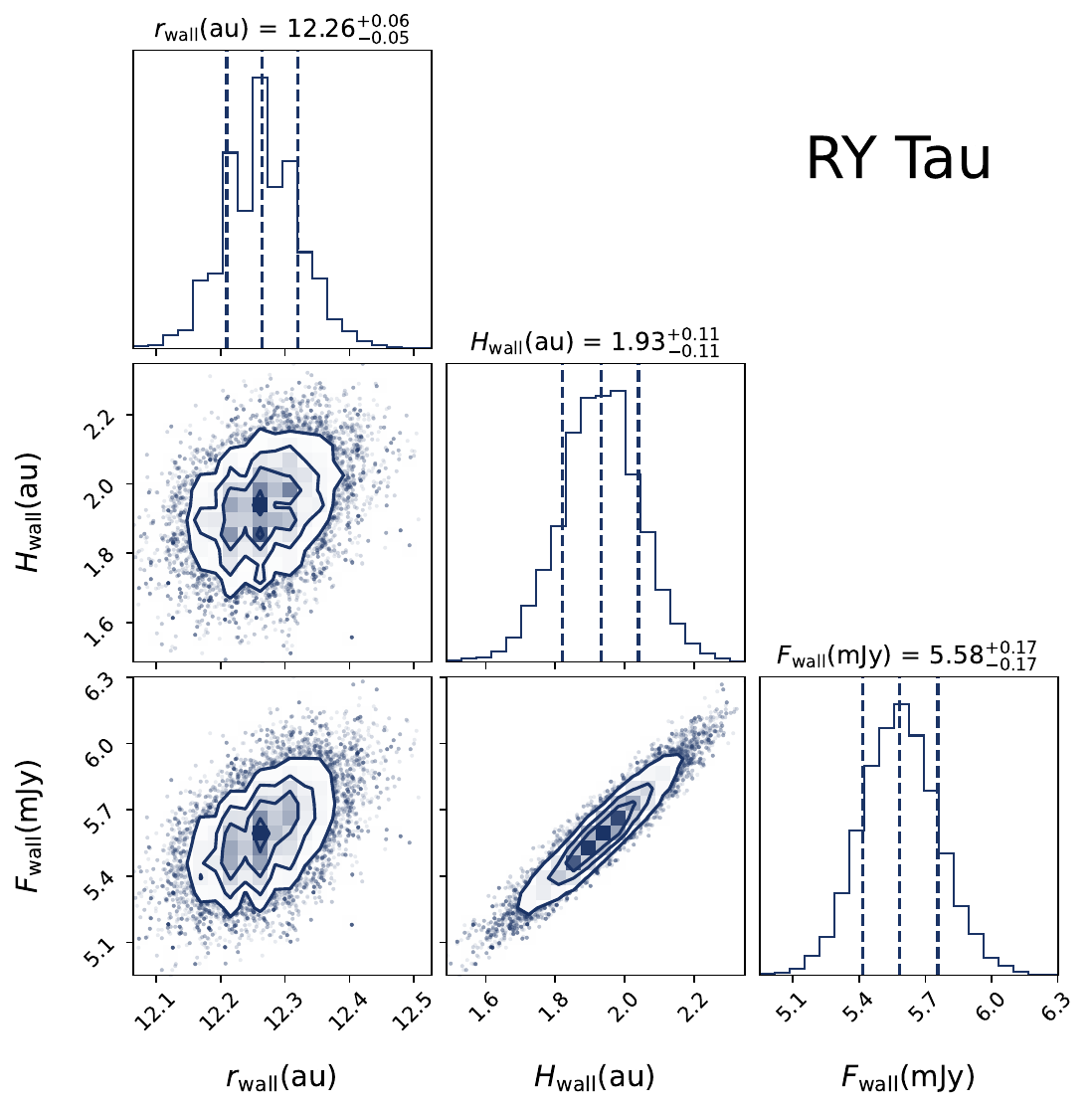}
    \caption{Cornerplots showing the posterior distributions and 2D correlations of wall-related parameters for CIDA 9 (top) and RY~Tau (bottom). The plots were computed using the last 1000 steps of the MCMC chain after convergence.}
    \label{fig:cornerplots}
\end{figure*}

We chose uniform priors with conservative ranges for $r_{\rm wall}$, $H_{\rm wall}$, $F_{\rm wall}$, $F_{\rm ring}$, and $\sigma_{\rm ring}$. For the remaining parameters, we used broad Gaussian priors centred at the values obtained from the {\tt frank} fit, which were consistent with previous studies in all cases \citep[e.g.,][]{Long2018}.

We also adopted a Gaussian likelihood function $\mathcal{L}$. For a given set of model parameters $\theta$, $\log(\mathcal{L}$) is then given by:
\begin{equation}
    \log \mathcal{L}({V_{\rm obs}} | \theta) = -\frac{1}{2} \bigl[\chi_{\rm Re}^2 + \chi_{\rm Im}^2\bigr] + C
\end{equation}

where 
\begin{equation}
    \chi_{\rm Re}^2 = \sum_{k=1}^{N} [\Re(V_{{\rm obs},k}) - Re(V_{{\rm model},k})]^2 w_k,
\end{equation}

and 
\begin{equation}
    \chi_{\rm Im}^2 = \sum_{k=1}^{N} [Im(V_{{\rm obs},k}) - Im(V_{{\rm model},k})]^2 w_k.
\end{equation}

Here, $V_{{\rm obs},k}$ and $V_{{\rm model},k}$ correspond to the $k$-th visibility of the observations and model, respectively, $Re$ and $Im$ are their real and imaginary parts, and $w_k=1/\sigma_k^2$ is the weight of the observed $k$-th visibility (the constant $C$ does not affect the MCMC calculation).

As a summary, the fitting process for each source is as follows:
\begin{enumerate}
    \item Determine the pixel size and number of pixels required for {\tt galario} based on the observed visibilities.
    \item Determine an initial {\tt frank} radial profile from the observations. This initial profile is not recalculated in each MCMC iteration (i.e., {\tt frank} is not ran multiple times during the fitting), but the profile is modified accordingly in the next steps.
    \item For a given set of parameters $\theta$, calculate the wall emission and correct the {\tt frank} profile as explained in Sec.~\ref{sec:model_description}. Then combine the disc and wall emission to produce a model image.
    \item Calculate the visibilities corresponding to this model using {\tt galario}.
    \item Compare the model visibilities with the observed ones to evaluate the posterior of $\theta$.
    \item Sample the posterior distribution of the model parameters with {\tt emcee} by repeating steps 3-5 until convergence is reached.
\end{enumerate}

We used 40 walkers for {\tt emcee}, and we ran the chains until their length was $>$100\,times the autocorrelation time (which is $\sim$ 40-60 iterations) to ensure that the MCMC process had converged. After this criterion was reached, the posterior distributions were computed by running another 1000\,additional iterations and discarding the previous steps as the burn-in phase.

\subsubsection{Modelling results}\label{sec:modelling_results}

The posterior distributions of both discs appear largely Gaussian for all parameters (except for $H_{\rm wall}$ in CIDA~9, for which only an upper limit can be obtained). Figure~\ref{fig:cornerplots} shows the results for the wall parameters $r_{\rm wall}$, $H_{\rm wall}$, and $F_{\rm wall}$, and Table~\ref{tab:fitting_results} lists the results for all the parameters. The geometric parameters are all compatible with those obtained from the {\tt frank} fit.

\begin{table}
    \caption{Fitting results. Note that, since we performed self-calibration, $\Delta \alpha$ and $\Delta \delta$ represent only deviations from the adopted phase centre for each source. A positive (negative) inclination means that the wall is visible on the southern (northern) side of the disc.} \label{tab:fitting_results}
    \centering
    \begin{tabular}{lcc}
    \hline
    & CIDA~9 & RY~Tau\\
    \hline
    $r_{\rm wall}$ (au) & $37.4\pm0.3$ & $12.26_{-0.05}^{+0.06}$\\
    $H_{\rm wall}$ (au) & $<$2.3 & $1.93\pm0.11$\\
    $F_{\rm wall}$ (mJy) & $3.3_{-0.3}^{+0.5}$ & $5.6\pm0.2$\\
    $F_{\rm ring}$ (mJy) & $4.7_{-0.3}^{+0.4}$ & $4.4\pm0.1$\\
    $\sigma_{\rm ring}$ (au) & $7.0\pm0.5$ & $0.94_{-0.08}^{+0.09}$\\
    $i$ (deg) & $-47.5\pm-0.1$ & $65.23\pm-0.01$\\
    $P.A.$ (deg) & $103.4\pm0.2$ & $23.11\pm0.02$\\
    $\Delta \alpha$ (mas) & $5.5\pm0.4$ & $0.7\pm0.1$\\
    $\Delta \delta$ (mas) & $-7.0\pm0.4$ & $20.8\pm0.1$\\
    Scaling $A$ (-) & $1.07\pm0.01$ & $0.995\pm0.001$\\
    \hline
    \end{tabular}
    \end{table}

In the case of CIDA~9, the cavity radius is $r_{\rm wall}=37.4\pm0.3\,{\rm au}$ and the fitting can only place an upper limit of $H_{\rm wall}<2.3\,{\rm au}$ to the thickness of the vertical rim (this value corresponds to the 84\,\% confidence interval). The total flux received from the wall at 1.3\,mm is $F_{\rm wall}=3.3_{-0.3}^{+0.5}\,{\rm mJy}$, which is $\sim$9\,\% of the total flux observed at 1.3\,mm. 

In contrast, the results for RY~Tau present a more restricting scenario, and the higher angular resolution of the observations constrain both the minimum and maximum vertical height of the wall. For this source, we obtain a cavity radius is $r_{\rm wall}=12.26_{-0.05}^{+0.06}\,{\rm au}$ and the wall vertical height is $H_{\rm wall}=1.93\pm0.11\,{\rm au}$. The wall flux is $F_{\rm wall}=5.6\pm0.2\,{\rm mJy}$, which in this case corresponds to 2.6\,\% of the total flux at 1.3\,mm.

To show the improvement of the geometric modelling compared to {\tt frank}, we imaged the residual visibilities of the {\tt frank} radial profiles and the best fit geometric models (i.e., the model with the highest likelihood value in the MCMC chain for each source). After subtracting the model visibilities from the observed ones, the residual visibilities were imaged in CASA using a {\tt robust} value of 0.5. The results are presented in Figs.~\ref{fig:residuals_CIDA9} and~\ref{fig:residuals_RY_Tau} and show a clear reduction in both the residual strength and structure when accounting for the wall emission. In the case of CIDA~9, the {\tt frank} fit yields positive residuals (up to 8\,$\sigma$) on the southern part of the disc where the wall emission is located, as well as negative residuals (up to -5\,$\sigma$) on the opposite side of the disc. When the wall is included, the residuals only show a localised 3-5\,$\sigma$ level on the South-East side. The improvement is more obvious for RY~Tau, which displays {\tt frank} residual levels up to 27\,$\sigma$ on the West side of the inner disc and negative -19\,$\sigma$ on the opposite side. Including the contribution of inner rim reduces the spatial extent and strength of the residuals, now ranging from 12 and -11\,$\sigma$. In both cases, the Bayesian Information Criterion (BIC) shows conclusive preference for the models including wall emission ($\Delta{\rm BIC}> 1000$).

\subsection{Radiative transfer modelling}\label{sec:modelling_mcfost}

\begin{figure*}
    \centering
	\includegraphics[width=0.96\hsize]{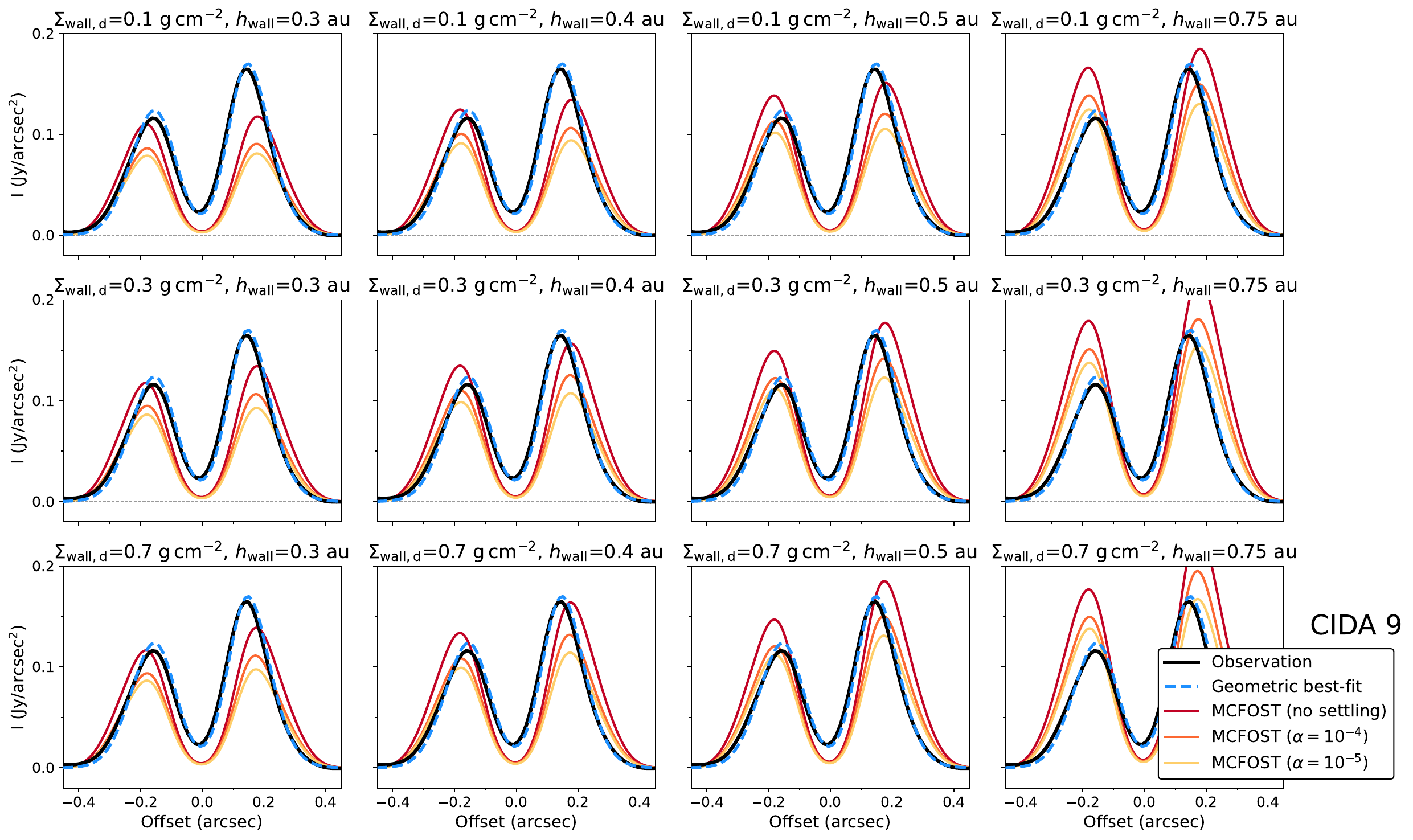}\
    \caption{Intensity cuts along the minor axis of the observation, best-fit geometric model, and the {\tt MCFOST} grid for CIDA~9. Models were first convolved with the same beam of the observations ($0.13\arcsec\times0.10\arcsec$, ${\rm PA}=2^{\circ}$). The observations are shown as a black line, the geometric model as a blue line, and three different settling scenarios are shown as a red (no settling), orange ($\alpha=10^{-4}$), and yellow ($\alpha=10^{-4}$). The title of each panel indicates the corresponding {\tt MCFOST} model parameters, where $\Sigma_{\rm wall, d}$ and $h_{\rm wall}$ are evaluated at 37.4\,au. The far side of the disc (where the disc wall is visible) corresponds to positive offsets.}
    \label{fig:CIDA9_MCFOST}
\end{figure*}

\begin{figure*}
    \centering
    \includegraphics[width=0.96\hsize]{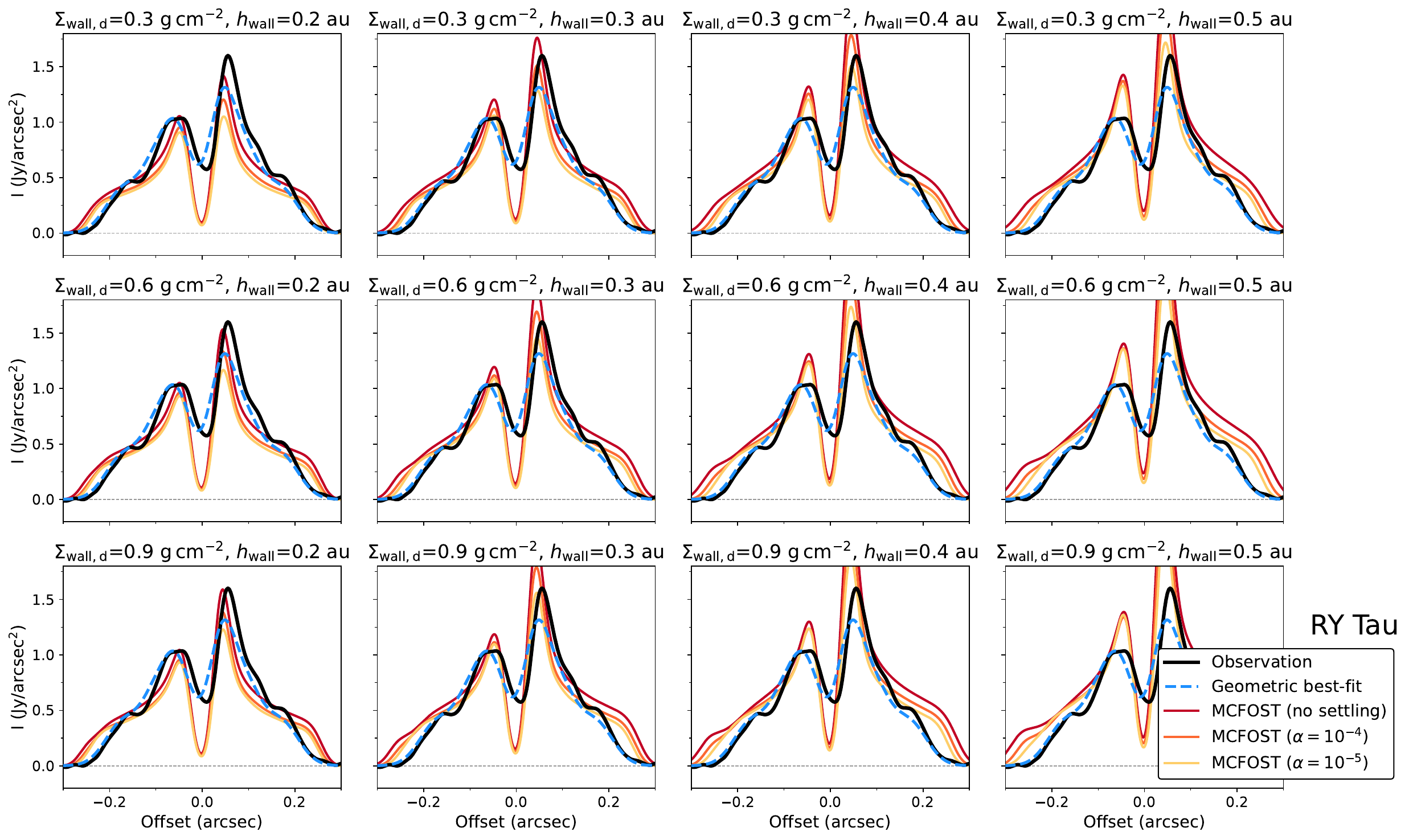}
    \caption{Intensity cuts along the minor axis of the observation, best-fit geometric model, and the {\tt MCFOST} grid for RY~Tau. Models were convolved with the the beam of the observations ($0.06\arcsec\times0.03\arcsec$, ${\rm PA}=27^{\circ}$). Colours, symbols, labels and orientation are as in Fig.~\ref{fig:CIDA9_MCFOST}. $\Sigma_{\rm wall, d}$ and $h_{\rm wall}$ are evaluated at 12.3\,au.}
    \label{fig:RY_Tau_MCFOST}
\end{figure*}

While the geometric model used above provides a fast and straightforward method to constraint $H_{\rm wall}$ from (sub)mm observations, the physical interpretation of this parameter is not the dust scale height $h_{\rm wall}$ at the cavity's location; instead, it is \emph{the vertical thickness of the inner rim which emits at an increased temperature as a result of being directly exposed to the stellar irradiation}. The connection between these two quantities depends on a number of factors such as the temperature and optical depth of the exposed wall at different heights (and therefore the local density and dust composition), the level of dust settling, and the shape of the wall. In fact, although we have adopted a sharp vertical wall for simplicity in our geometric model, both theory and infrared interferometric observations suggest that the inner rim of discs is a complex environment that is likely curved and has some radial extent. For small cavities, the inner disc is further sculpted by dust sublimation \citep[e.g.,][]{Dullemond2001, Isella2005, Kama2009, Dullemond2010, Flock2017, Kluska2020}, while for larger cavities such as the ones of CIDA~9 and RY~Tau their shape will instead be determined by the mechanism carving the inner gap. In all cases, assuming a sharp, vertically isothermal wall is still a crude approximation.

To get an estimate of the dust scale height of these discs, we used the radiative transfer code {\tt MCFOST} \citep{MCFOST,MCFOST2}. These models produce more realistic representations of the wall emission at the cost of longer computation times, and thus we decided not to run an MCMC analysis but to produce a grid of models for each source instead. As with the geometric models, we are not interested in the overall structure of the discs and our main aim is to determine which values of $h_{\rm wall}$ would be compatible with the observed asymmetries. Therefore, we used a power-law profile for the surface density of the discs $\Sigma(r) \propto r^{-1/2}$ extending from the $r_{\rm wall}$ derived in Sec.~\ref{sec:modelling_results} (37.4 and 12.3\,au for CIDA~9 and RY~Tau, respectively) to 80\,au, which is an appropriate dust outer radius in both cases based on the {\tt frank} radial profiles. The disc flaring was set to $h(r) \propto r^{1.125}$, and we adopted a gas-to-dust mass ratio of 100. For the dust properties, we used a grain size distribution following d$n(a) \propto a^{-3.5} $d$a$ from $a_{\rm min}=0.03\,\mu{\rm m}$ to $a_{\rm max}=3\,{\rm mm}$, the DIANA dust composition \citep{Woitke2016}, and we used the Distribution of Hollow Spheres method \citep{Min2016} to calculate the dust scattering and absorption. The adopted stellar parameters and distances are as indicated in Sec.~\ref{sec:sample_and_observations} ($T=3590$\,K, $M_*=0.6\,M_\odot$, $R_*=1.18\,R_\odot$, and $d$=175\,pc for CIDA~9, and $T=5945\,{\rm K}$, $M_*=1.95\,M_\odot$, $R_*=3.25\,R_\odot$ and $d=138\,{\rm pc}$ for RY~Tau).

We focused on the effect of three parameters in our model grids: the dust surface density $\Sigma_{\rm wall, d}$ and (gas) scale height $h_{\rm wall}$ at the location of the wall, and the level of dust settling. For $\Sigma_{\rm wall, d}$, we first converted the 1.3\,mm flux into a dust mass assuming optically thin emission, a dust opacity at 1.3\,mm of $\kappa_{\rm 1.3\,mm}=2.3\,{\rm cm^{2}\,g^{-1}}$ and a dust temperature of 20\,K \citep[e.g.,][]{Andrews2005}. This results in dust masses of $9.6\times 10^{-5}M_\odot$ for CIDA~9 and $3.5\times 10^{-4}M_\odot$ for RY~Tau which, given the adopted surface density profiles and inner and outer radii, correspond to dust surface densities of 0.07\,${\rm g\,cm^{-2}}$ for CIDA~9 and 0.3\,${\rm g\,cm^{-2}}$ for RY~Tau at the location of their inner radii. For CIDA~9, we produced models with 0.5, 1, 5, and 10 times its nominal surface density, while for RY~Tau we calculated a few additional cases including 0.5, 1, 2, 3, and 5 times its nominal value. For the scale height at the location of the inner wall $h_{\rm wall}$, we tested 0.1, 0.2, 0.3, 0.4, 0.5, 0.75, and 1\,au for both sources. Dust settling was explored by assuming three scenarios: no settling (i.e., a similar scale height for the gas and all dust particles), and two different levels of settling using the \citet{Fromang2009} prescription (see their Eq. 19), which results in a different vertical distribution for each particle size (with larger grains more concentrated in the midplane). In this case, settling is controlled through the gas turbulence parameter $\alpha$ introduced by \citet{Shakura1973}, and we tried $\alpha$ values of $10^{-4}$ and $10^{-5}$ which result in significantly settled discs and match observational estimates \citep[e.g.,][]{Pinte2016,Flaherty2015,Flaherty2017,Villenave2022}.

The {\tt MCFOST} models were then confronted with the observations and the best-fit geometric model from Sec.~\ref{sec:modelling_results}. For this purpose, we first convolved the geometric and {\tt MCFOST} models of each disc with the beam size of their {\tt robust=0.5} synthesised images ($0.13\arcsec\times0.10\arcsec$, ${\rm PA}=2^{\circ}$ for CIDA~9, and $0.06\arcsec\times0.035\arcsec$, ${\rm PA}=27^{\circ}$ for RY~Tau), and then compared their intensity profiles along the minor axis. The results are presented in Fig.~\ref{fig:CIDA9_MCFOST} for CIDA~9, and Fig.~\ref{fig:RY_Tau_MCFOST} for RY~Tau (for the sake of clarity, we do not show the complete grid of models in each case, but focus on $\Sigma_{\rm wall, d}-h_{\rm wall}$ values around good fits to the observations). 

The best-fit geometric model of CIDA~9 closely follows the observations, but only {\tt MCFOST} models with high surface densities provide sufficient contrast between the near and far-sides of the disc. For the case without dust settling, we find reasonable matches using $\Sigma_{\rm wall, d}$ values of 0.3-0.5 ${\rm g\,cm^{-2}}$ and a scale height of $h_{\rm wall}=0.4\,{\rm au}$. Alternatively, we can also reproduce the overall shape and intensity of the asymmetry using a settled disc ($\alpha=10^{-4}$) with $\Sigma_{\rm wall, d}=0.5\,{\rm g\,cm^{-2}}$ and a scale height $h_{\rm wall}=0.5\,{\rm au}$. These models have midplane dust densities of $2.3-5.5\times10^{-14}\,{\rm g\,cm^{-3}}$, and the settled case ($\alpha=10^{-4}$) has a gas-to-dust mass ratio in the midplane of $\sim$65 (although these values are highly uncertain due to the several degeneracies and simplifications of our modelling, see Sec.~\ref{sec:caveats}). It is worth mentioning that a high dust surface density (5-10 times higher than the value based on dust mass from the millimetre flux) is required to make the inner regions of the disc sufficiently optically thick, causing the wall to be visible on the far-side only and creating the observed contrast between both sides. This does not necessarily imply a much higher disc mass, but just a high optical depth near the disc cavity.

The best-fit geometric model for RY~Tau also resembles the observed intensity profile, although the modelled wall emission appears slightly fainter than in the data. We find a better agreement for some of the {\tt MCFOST} models, with the best results depending on the assumed settling: a scale height of $h_{\rm wall}=0.2\,{\rm au}$ and a dust surface density of $\Sigma_{\rm wall, d}=0.9\,{\rm g\,cm^{-2}}$ for the case without settling, a scale height of $h_{\rm wall}=0.3\,{\rm au}$ for $\Sigma_{\rm wall, d}=0.6\,{\rm g\,cm^{-2}}$ and $\alpha=10^{-4}$, and a similar scale height for $\Sigma_{\rm wall, d}=0.9\,{\rm g\,cm^{-2}}$ and $\alpha=10^{-5}$. The midplane dust density of these models at the location of the cavity ranges from $8.6\times10^{-14}-2.3\times10^{-13}\,{\rm g\,cm^{-3}}$, and the corresponding gas-to-dust ratios go from 100 (no settling) to 35 for the very settled case ($\alpha=10^{-5}$). We note that the central cavity of the {\tt MCFOST} profiles is deeper than in the observations and the geometric fit, indicating that it is not completely devoid of dust \citep[as also suggested by near-IR interferometry,][]{Davies2020}.

\subsection{Comparison between the geometric and {\tt MCFOST} models}\label{sec:model_comparison}

Since $H_{\rm wall}$ and $h_{\rm wall}$ are not directly comparable, we checked the consistency of both modeling approaches by taking the {\tt MCFOST} best-fit image of RY~Tau ($\Sigma_{\rm wall, d}=0.9\,{\rm g\,cm^{-2}}$, $h$=0.2\,au, no settling) and applying our geometric model to confirm that we obtain a similar $H_{\rm wall}\sim$2\,au value. For this purpose, we first extracted the visibilities of the {\tt MCFOST} image for the same $uv$-coverage of the observations using {\tt galario}. We obtained the radial profile of those visibilities with {\tt frank}, and then modeled the visibilities following the procedure in Sec.~\ref{sec:modelling_process}. This test yielded two main results: first, the derived location of the inner disc is slightly smaller than the true value used in {\tt MCFOST} (10.38$_{-0.07}^{+0.02}$\,au vs 12.26\,au in the {\tt MCFOST} model). This is not surprising given the different treatment and shape of the wall emission in MCFOST compared to our simple geometric prescription. This does not affect our analysis, but implies that the cavity radii derived with the geometric models may be underestimated. The second result is that we obtain $H_{\rm wall}=2.21_{-0.10}^{+0.09}$\,au, very similar to the measurement of $H_{\rm wall}=1.93\pm0.1$\,au from the observations. We note that the best {\tt MCFOST} case in Sec.~\ref{sec:modelling_mcfost} was determined from a coarse grid and does not represent a detailed fit to the observations, so the small difference between both $H_{\rm wall}$ values is to be expected and we find both approaches to be compatible.

\subsection{Implications for CIDA~9 and RY~Tau}\label{sec:discussion_objects}

If we assume that the crescent-shaped asymmetries in CIDA~9 and RY~Tau are due to their exposed inner rims, then our analysis provides information about the vertical extent of these walls. The geometric modelling placed an upper of limit of $H_{\rm wall}<2.3\,{\rm au}$ for CIDA~9 and estimated it to be $1.93\pm0.11\,{\rm au}$ for RY~Tau. The {\tt MCFOST} models also allowed us to explore the corresponding dust scale heights at the inner rim: for CIDA~9, the {\tt MCFOST} models imply a dust scale height of $\sim0.4\,{\rm au}$, while we find $h_{\rm d}=0.2\,{\rm au}$ for RY~Tau. Combined with the flaring index of 1.125 used in the {\tt MCFOST} models, these yield dust scale heights of $\sim 1-2\,{\rm au}$ at 100\,au for these sources, comparable to the values obtained for the edge-on disc Oph163131 \citep[$<1\,{\rm au}$,][]{Villenave2022}, for HL~Tau \citep[$<2\,{\rm au}$][]{Pinte2016}, and the $\lesssim 4\,{\rm au}$ constraints derived for six discs in \citet{Pizzati2023}. The ratio between the dust scale height from {\tt MCFOST} and the disc thickness from the geometric modelling is $h_{\rm d}/H_{\rm wall}\sim0.1$ for RY~Tau (and a lower limit of $>0.08$ for CIDA~9), indicating that the emission from the wall is still optically thick several scale heights above the midplane when measured at millimetre wavelengths. Additionally, despite the range of settling levels that could reproduce the observations, the midplane dust densities of the best models remain similar (within a factor of 2-3) for each source ($\sim5\times10^{-14}\,{\rm g\,cm^{-3}}$ for CIDA~9 and $\sim10^{-13}\,{\rm g\,cm^{-3}}$in the case of RY~Tau).

%%%%%%%%%%%%%%%%%%%%%%%%%%%%%%%%%%%%%%%%%%%%%%%%%%%%%%%%%%%%%
% DISCUSSION
\section{Crescent-shaped emission at millimetre wavelengths interpreted as inner disc rims}\label{sec:discussion}

\begin{figure*}
    \centering  
	\includegraphics[width=\hsize]{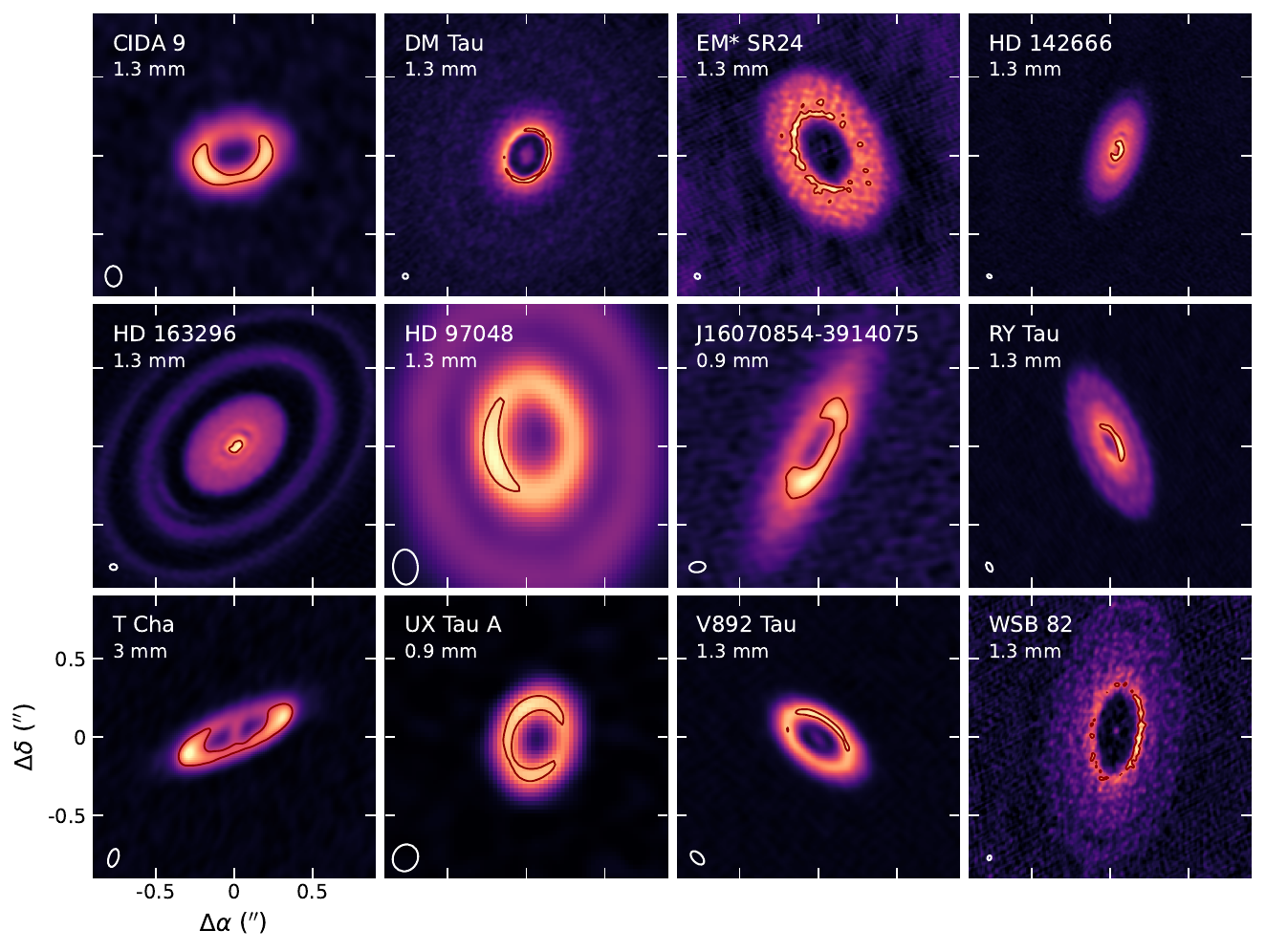}    
    \caption{Examples of sources with asymmetries compatible with inner walls. All images show the 1.3\,mm dust continuum except for UX~Tau~A and J16070854-3914075 (0.88~mm), and T~Cha (3~mm). Orientation information from scattered light and/or gas observations is available for all objects except three (CIDA~9, J16070854-3914075 and WSB~82). In all those cases, the orientation implies that the crescent-shaped asymmetries are located on the far side where the exposed wall is expected. Contour levels are shown to emphasise the asymmetries.}
    \label{fig:walls_gallery}
\end{figure*}

If crescent-shaped asymmetries in discs at millimetre wavelengths arise from their hot, exposed inner rims, then we can ask ourselves two related questions: \textit{what conditions facilitate the detection of such asymmetries?}, and \textit{are disc walls regularly detected?}. 

On the first point, for the asymmetry to appear the emission near the cavity radius should be optically thick so that the wall is only visible on the far side. Evidence for (at least partially) optically thick emission from protoplanetary discs at millimetre wavelengths has become increasingly stronger in recent years through different studies, including possibly underestimated disc dust masses \citep{Ballering2019,Ribas2020,Xin2023,Rilinger2023}, the identification of a disc radius-luminosity relation at millimetre wavelengths \citep{Tripathi2017,Andrews2018}, and multiwavelength observations of various targets \citep[e.g.,][]{Huang2018,Macias2021,Sierra2021,Guidi2022,Ribas2023}. It is then plausible that this requirement is met in a significant number of discs. The condition of optically thick emission also implies that the asymmetry contrast will decrease (or disappear) at increasing wavelengths as optical depth decreases – multiwavelength observations may therefore be used to test if an azimuthal asymmetry along the disc minor axis is tracing a true overdensity in the dust, or if it may be due to an inner rim instead.

Furthermore, if we consider the wall to be optically thick, then its emission increases for geometrically thicker discs and/or larger cavities (which increases the wall's emitting area), discs seen with higher inclination (which increases the wall's solid angle, as long as it is not blocked by the disc itself), as well as for hotter walls (which increases the wall's intensity). Thus, walls should be easier to detect in discs seen with moderate-large inclinations and with sufficiently wide cavities for the wall's solid angle to be large, but at disc radii where the emission is still optically thick.

If the conditions to detect the emission from walls are common, crescent-shaped asymmetries aligned with the disc minor axis may also be common. To investigate this, we inspected published intermediate/high-angular resolution ALMA observations and searched for such asymmetries. Figure~\ref{fig:walls_gallery} shows twelve examples for which we could identify asymmetries resembling the emission from walls, including CIDA~9, DM~Tau, EM*~SR24, HD~142666, HD~163296, HD~97048, 2MASS~J16070854-3914075, RY~Tau, T~Cha, UX~Tau~A, V892~Tau, and WSB~82 \citep[][]{Long2018,Andrews2018,Cieza2021,Hashimoto2021,Alaguero2024}\footnote{ALMA projects 2015.1.00888.S, 2015.1.00979.S, 2016.1.00484.L, 2016.1.01164.S, 2017.1.01460.S, 2018.1.00028.S, 2019.1.01091.S, 2021.1.01137.S, and 2022.1.00742.S.}. This list is not exhaustive and, since we identified the sources by eye, it is likely that several other examples exist already in the ALMA archive. The majority of these data were taken at 1.3\,mm, but two of them were observed at 0.9\,mm (2MASS~J16070854-3914075 and UX~Tau~A) and there is even one example at 3\,mm (T~Cha).

These sources serve to further test the hypothesis that crescent-shaped asymmetries at millimetre wavelengths may arise from exposed inner rims: in that case, the asymmetries should appear on the far side of the disc. We found information about the orientation of nine out of the twelve systems shown in Fig.~\ref{fig:walls_gallery}, both from scattered-light/near-IR interferometry \citep[EM*~SR24, HD~142666, HD~163296, HD~97048, RY~Tau, T~Cha, UX~Tau~A,][]{Ginski2016,Rich2019,Menard2020,Garufi2022,Valegard2022,Weber2023} and the shape of the gas emission traced by ALMA \citep[DM~Tau, HD~163296, V892~Tau,][as well as archival observations]{Teague2019, Alaguero2024}. The crescent-shaped asymmetry indeed appears on the far side for each of the nine discs with orientation information, amounting to a $<$0.2\,\% chance of this being a random phenomenon. This strongly suggest that these asymmetries are geometric in nature (at least in a non-negligible number of cases) and they may be used to identify the near/far sides of the disc from millimetre observations. This effect can also appear in the walls of rings external to the inner disk if the scale height and inclination are sufficiently high, as recently reported in HL~Tau \citep{GuerraAlvarado2024}. More importantly, the modelling of these asymmetries can offer both new and complementary insights about the vertical structure of protoplanetary discs, and could also be used to test (or in combination with) results from other methodologies.

\subsection{Caveats and alternative interpretations}\label{sec:caveats}

While here we interpret crescent-shaped asymmetries in discs as signatures of their inner rims, azimuthal asymmetries may instead reflect true overdensities induced by dust trapping in vortices \citep[e.g.,][]{Godon2000,Meheut2012,Baruteau2016} or by the effect of massive companions \citep[e.g.,][]{Ragusa2017,Dong2018_MWC758}. Several sources have vortex-like asymmetries that are likely incompatible with originating in the disc wall, since they are not aligned with the disc minor axis \citep[e.g., HD~142527, Oph~IRS~48, LkH$\alpha$~330, AB~Aur, HD~34282, HD~135344B,][]{Tang2012,Fukagawa2013,vanderMarel2013,vanderPlas2017,Cazzoletti2018,Pinilla2022}. As an example, \citet{Soon2017} presented detailed modelling of the disc around HD~142527 using both optically thick and thin models, showing that the former resulted in crescent-shape emission from the wall that was geometrically incompatible with the observations. However, in the presence of an inclined inner disc, it is possible for the asymmetry to appear on the far side but not aligned with the minor axis due to the shadow casted by the inner disc, as proposed for SY~Cha by \citet{Orihara2023}. On the other hand, even asymmetries where the alignment appears to be correct may still be due to vortices, and both explanations are not mutually exclusive: for instance, \citet{Harsono2024} attributed the shape of the millimetre emission in CIDA~9 to a vortex and found tentative evidence for an azimuthal shift of the emission peaks between the 1.3\,mm and 3\,mm observations. The observed contrast between the North-South sides of the disc is $\sim$30\,\% both at 1.3\,mm and 3\,mm, which would imply either a real overdensity or optically thick emission\footnote{Interestingly, \citet{Harsono2024} determined the spectral index to CIDA~9 between these wavelengths to be 2-3 based on a pixel-by-pixel comparison and joint imaging of the 1.3\,mm and 3\,mm data using the {\tt mtmfs} deconvolver \citep{Rau2011}. However, their results depend on the number of adopted spectral terms. When using $n=2$ (the same number of available observational constraints) they obtain $\alpha\approx2$, compatible with optically thick emission.}. In fact, the residuals of CIDA~9 in Fig.~\ref{fig:residuals_CIDA9} still show localised emission toward the East, a feature that is also visible at 3\,mm \citep{Harsono2024} and may trace a local increase in the dust density.  Similarly, \citet{Hashimoto2021} reported two peaks on the far side of DM~Tau, and hypothesise that they may also be due to two vortices or the destruction of a bigger one, among other explanations (we note that the best-fit model they obtain for the main blob in the disc closely resembles the expected shape of a potential wall). For these cases, no conclusive evidence exists regarding the true nature of the asymmetries. Additional dust observations at multiple wavelengths, combined with high angular resolution data tracing the gas kinematics, are needed to confirm their origin.

Likewise, we also identified two sources with strong asymmetries along their minor axis, namely RY~Lup \citep{Francis2020} and IRC~101 \citep{vanTerwisga2020}. These two discs, however, appear morphologically distinct from the others in Fig.~\ref{fig:walls_gallery}: they host very large cavities (70\,au and 300\,au, respectively), are significantly inclined ($>$60\,deg), and display very strong contrasts between the intensity at their minor axes. It is unlikely that such extreme cases can be reproduced by disc inner rims. In fact, although no orientation information exists for IRC~101, the modelling of SPHERE observations of RY~Lup presented in \citet{Langlois2018} indicates that the millimetre asymmetry is located in the near side for this disc, and is thus incompatible with wall emission as studied here.

It is also worth noticing that some of the discs shown in Fig.~\ref{fig:walls_gallery} also display two bright peaks along their major axes, including J16070854, T~Cha, and even CIDA~9. While this may be due to the intermediate resolution of the observations and the orientation/elongation of the beam, such effect is also visible in optically thin, geometrically thick discs due to limb brightening \citep{Doi2021}. Both effects (crescent-shaped asymmetries due to inner walls and limb brightening) appear under different conditions (optically thick/thin emission) and therefore will not be visible in the same ring. However, if a disc contains multiple rings with different optical depths, it would be possible to observe both effects in the same system (i.e., a wall asymmetry and an increase in brightness at the disc ansae). One promising case of such scenario is CIDA~9, where the {\tt frank} radial profile suggests two rings (see the axisymmetric model in Fig.~\ref{fig:residuals_CIDA9}) – an optically thick inner ring and optically thin, geometrically thick outer one could explain the appearance of the disc at the available angular resolutions.

Finally, we emphasise that the derived scale heights depend on the correctness of the adopted model and parameters, and should be considered as an approximation. None of the models used in this work account for any physical structure of the rim and, instead, they assume a sharp wall. In reality, the shape of the rim will be determined by the mechanism producing the cavity, as well as by the local disc conditions. Similarly, the adopted stellar parameters and disc structure may also not be accurate. In particular, the stellar parameters for RY~Tau are quite uncertain, and near-IR interferometric observations also suggest the presence of some material in its inner regions \citep{Davies2020}. Evidence for this is also found in the ALMA images, where the intensity in the central gap does not appear to reach the zero level. In that case, it is possible that the true wall temperature of RY~Tau is lower than the one derived with {\tt MCFOST}, resulting in an overestimation of the true wall intensity and thus an underestimation of its dust scale height in our work.

%%%%%%%%%%%%%%%%%%%%%%%%%%%%%%%%%%%%%%%%%%%%%%%%%%%%%%%%%%%%%
% CONCLUSSIONS
\section{Conclusions}

In this paper we interpreted crescent-shaped asymmetries seen in millimetre observations of protoplanetary discs as emission from their hot inner rims. Such asymmetries could arise in discs with cavities seen at moderate/high inclinations if the emission near the gap is optically thick. This exposes the rim on the far side of the disc only, resulting in an azimuthal asymmetry similar to the ones observed in a number of discs.

We modelled these asymmetries in the ALMA 1.3\,mm observations of CIDA~9 and RY~Tau using two different approaches, first with a simple geometric prescription for the wall emission, and then using more accurate radiative transfer models. Our main results are:
\begin{itemize}
\item Including the emission from the wall yields significantly better residuals than axisymmetric models, with little to no structure left in them.
\item The geometric modelling places an upper limit on the vertical thickness of the wall $H_d$ for CIDA~9 ($<2.3\,{\rm au}$ at $r=37\,{\rm au}$) and constrains it for RY~Tau ($1.93\pm0.11\,{\rm au}$ at $r=12\,{\rm au}$).
\item By comparing the intensity profiles along the minor axis, we also estimate the dust scale heights $h_d$ of these two sources at their walls location ($\sim0.4\,{\rm au}$ and $\sim0.2\,{\rm au}$ for CIDA~9 and RY~Tau, respectively). Good matches to the observations also exist for more settled models ($\alpha=10^{-4}-10^{-5}$) and higher dust surface densities. These models have midplane dust densities of $\sim5\times10^{-14}\,{\rm g\,cm^{-3}}$ for CIDA~9 and $\sim10^{-13}\,{\rm g\,cm^{-3}}$ for RY~Tau, and midplane 
gas-to-dust ratios between 30-100.

\end{itemize}

We also identified a total of twelve discs with crescent-shaped asymmetries based on visual inspection of published ALMA observations. The orientation is known for nine of these systems, placing the millimetre asymmetry on the far side of the disc in all cases. This suggests a geometric origin for the asymmetries and lends support to the idea that they may arise from exposed inner walls, but high-resolution, multiwavelength continuum data and/or gas kinematic information are needed to investigate if they trace other phenomena (such as vortices) instead. If these asymmetries really probe the dust scale height of discs, then the methodology presented here provides a complementary approach to study their vertical structure.

\section*{Acknowledgements}

We thank the referee, Takayuki Muto, for his review and comments which helped to improve the quality of this manuscript. We are grateful to Feng Long for sharing the self-calibrated visibilities of the compact configuration of RY Tau, François Ménard for the ALMA observations of SSTc2d J160708, and Jun Hashimoto for the DM Tau data. We also thank Nicolás Cuello, Pietro Curone, Cristiano Longarini, Enrique Macías, and François Ménard for various insightful discussions. AR and CJC have been supported by the UK Science and Technology Facilities Council (STFC) via the consolidated grant ST/W000997/1. AR, CJC, and FZ have also been supported by the European Union’s Horizon 2020 research and innovation programme under the Marie Sklodowska-Curie grant agreement No. 823823 (RISE DUSTBUSTERS project). This paper makes use of the following ALMA data: ADS/JAO.ALMA\#2015.1.00888.S, 2015.1.00979.S, 2016.1.00484.L, 2016.1.01164.S, 2017.1.01460.S, 2018.1.00028.S, 2019.1.01091.S, 2021.1.01137.S, and 2022.1.00742.S. ALMA is a partnership of ESO (representing its member states), NSF (USA) and NINS (Japan), together with NRC (Canada), MOST and ASIAA (Taiwan), and KASI (Republic of Korea), in cooperation with the Republic of Chile. The Joint ALMA Observatory is operated by ESO, AUI/NRAO and NAOJ.

%%%%%%%%%%%%%%%%%%%%%%%%%%%%%%%%%%%%%%%%%%%%%%%%%%
\section*{Data Availability}

All the observations used in this work are publicly available thought the ALMA archive.


\bibliographystyle{mnras}
\bibliography{biblio} % if your bibtex file is called example.bib

\begin{thebibliography}{}
\makeatletter
\relax
\def\mn@urlcharsother{\let\do\@makeother \do\$\do\&\do\#\do\^\do\_\do\%\do\~}
\def\mn@doi{\begingroup\mn@urlcharsother \@ifnextchar [ {\mn@doi@}
  {\mn@doi@[]}}
\def\mn@doi@[#1]#2{\def\@tempa{#1}\ifx\@tempa\@empty \href
  {http://dx.doi.org/#2} {doi:#2}\else \href {http://dx.doi.org/#2} {#1}\fi
  \endgroup}
\def\mn@eprint#1#2{\mn@eprint@#1:#2::\@nil}
\def\mn@eprint@arXiv#1{\href {http://arxiv.org/abs/#1} {{\tt arXiv:#1}}}
\def\mn@eprint@dblp#1{\href {http://dblp.uni-trier.de/rec/bibtex/#1.xml}
  {dblp:#1}}
\def\mn@eprint@#1:#2:#3:#4\@nil{\def\@tempa {#1}\def\@tempb {#2}\def\@tempc
  {#3}\ifx \@tempc \@empty \let \@tempc \@tempb \let \@tempb \@tempa \fi \ifx
  \@tempb \@empty \def\@tempb {arXiv}\fi \@ifundefined
  {mn@eprint@\@tempb}{\@tempb:\@tempc}{\expandafter \expandafter \csname
  mn@eprint@\@tempb\endcsname \expandafter{\@tempc}}}

\bibitem[\protect\citeauthoryear{{ALMA Partnership} et~al.,}{{ALMA Partnership}
  et~al.}{2015}]{HLTau}
{ALMA Partnership} et~al., 2015, \mn@doi [\apjl] {10.1088/2041-8205/808/1/L3},
  \href {http://adsabs.harvard.edu/abs/2015ApJ...808L...3A} {808, L3}

\bibitem[\protect\citeauthoryear{{Alaguero} et~al.,}{{Alaguero}
  et~al.}{2024}]{Alaguero2024}
{Alaguero} A.,  et~al., 2024, \mn@doi [arXiv e-prints]
  {10.48550/arXiv.2405.12593}, \href
  {https://ui.adsabs.harvard.edu/abs/2024arXiv240512593A} {p. arXiv:2405.12593}

\bibitem[\protect\citeauthoryear{{Andrews} \& {Williams}}{{Andrews} \&
  {Williams}}{2005}]{Andrews2005}
{Andrews} S.~M.,  {Williams} J.~P.,  2005, \mn@doi [\apj] {10.1086/432712},
  \href {http://adsabs.harvard.edu/abs/2005ApJ...631.1134A} {631, 1134}

\bibitem[\protect\citeauthoryear{{Andrews}, {Terrell}, {Tripathi}, {Ansdell},
  {Williams}  \& {Wilner}}{{Andrews} et~al.}{2018a}]{Andrews2018}
{Andrews} S.~M.,  {Terrell} M.,  {Tripathi} A.,  {Ansdell} M.,  {Williams}
  J.~P.,   {Wilner} D.~J.,  2018a, \mn@doi [\apj] {10.3847/1538-4357/aadd9f},
  \href {https://ui.adsabs.harvard.edu/abs/2018ApJ...865..157A} {865, 157}

\bibitem[\protect\citeauthoryear{{Andrews} et~al.,}{{Andrews}
  et~al.}{2018b}]{Andrews2018_DSHARP}
{Andrews} S.~M.,  et~al., 2018b, \mn@doi [\apjl] {10.3847/2041-8213/aaf741},
  \href {https://ui.adsabs.harvard.edu/abs/2018ApJ...869L..41A} {869, L41}

\bibitem[\protect\citeauthoryear{{Avenhaus} et~al.,}{{Avenhaus}
  et~al.}{2018}]{Avenhaus2018}
{Avenhaus} H.,  et~al., 2018, \mn@doi [\apj] {10.3847/1538-4357/aab846}, \href
  {https://ui.adsabs.harvard.edu/abs/2018ApJ...863...44A} {863, 44}

\bibitem[\protect\citeauthoryear{{Bae}, {Isella}, {Zhu}, {Martin}, {Okuzumi}
  \& {Suriano}}{{Bae} et~al.}{2023}]{Bae2023}
{Bae} J.,  {Isella} A.,  {Zhu} Z.,  {Martin} R.,  {Okuzumi} S.,   {Suriano} S.,
   2023, in {Inutsuka} S.,  {Aikawa} Y.,  {Muto} T.,  {Tomida} K.,   {Tamura}
  M.,  eds,  Astronomical Society of the Pacific Conference Series Vol. 534,
  Protostars and Planets VII. p.~423 (\mn@eprint {arXiv} {2210.13314}),
  \mn@doi{10.48550/arXiv.2210.13314}

\bibitem[\protect\citeauthoryear{{Ballering} \& {Eisner}}{{Ballering} \&
  {Eisner}}{2019}]{Ballering2019}
{Ballering} N.~P.,  {Eisner} J.~A.,  2019, \mn@doi [\aj]
  {10.3847/1538-3881/ab0a56}, \href
  {https://ui.adsabs.harvard.edu/abs/2019AJ....157..144B} {157, 144}

\bibitem[\protect\citeauthoryear{{Baruteau} \& {Zhu}}{{Baruteau} \&
  {Zhu}}{2016}]{Baruteau2016}
{Baruteau} C.,  {Zhu} Z.,  2016, \mn@doi [\mnras] {10.1093/mnras/stv2527},
  \href {https://ui.adsabs.harvard.edu/abs/2016MNRAS.458.3927B} {458, 3927}

\bibitem[\protect\citeauthoryear{{Benisty} et~al.,}{{Benisty}
  et~al.}{2023}]{Benisty2023}
{Benisty} M.,  et~al., 2023, in {Inutsuka} S.,  {Aikawa} Y.,  {Muto} T.,
  {Tomida} K.,   {Tamura} M.,  eds,  Astronomical Society of the Pacific
  Conference Series Vol. 534, Protostars and Planets VII. p.~605 (\mn@eprint
  {arXiv} {2203.09991}), \mn@doi{10.48550/arXiv.2203.09991}

\bibitem[\protect\citeauthoryear{{Boehler} et~al.,}{{Boehler}
  et~al.}{2021}]{Boehler2021}
{Boehler} Y.,  et~al., 2021, \mn@doi [\aap] {10.1051/0004-6361/202040089},
  \href {https://ui.adsabs.harvard.edu/abs/2021A&A...650A..59B} {650, A59}

\bibitem[\protect\citeauthoryear{{Calvet}, {Muzerolle}, {Brice{\~n}o},
  {Hern{\'a}ndez}, {Hartmann}, {Saucedo}  \& {Gordon}}{{Calvet}
  et~al.}{2004}]{Calvet2004}
{Calvet} N.,  {Muzerolle} J.,  {Brice{\~n}o} C.,  {Hern{\'a}ndez} J.,
  {Hartmann} L.,  {Saucedo} J.~L.,   {Gordon} K.~D.,  2004, \mn@doi [\aj]
  {10.1086/422733}, \href
  {https://ui.adsabs.harvard.edu/abs/2004AJ....128.1294C} {128, 1294}

\bibitem[\protect\citeauthoryear{Carrera, Simon, Li, Kretke  \& Klahr}{Carrera
  et~al.}{2021}]{Carrera2021}
Carrera D.,  Simon J.~B.,  Li R.,  Kretke K.~A.,   Klahr H.,  2021, \mn@doi
  [The Astronomical Journal] {10.3847/1538-3881/abd4d9}, 161, 96

\bibitem[\protect\citeauthoryear{{Cazzoletti} et~al.,}{{Cazzoletti}
  et~al.}{2018}]{Cazzoletti2018}
{Cazzoletti} P.,  et~al., 2018, \mn@doi [\aap] {10.1051/0004-6361/201834006},
  \href {https://ui.adsabs.harvard.edu/abs/2018A&A...619A.161C} {619, A161}

\bibitem[\protect\citeauthoryear{{Cieza} et~al.,}{{Cieza}
  et~al.}{2021}]{Cieza2021}
{Cieza} L.~A.,  et~al., 2021, \mn@doi [\mnras] {10.1093/mnras/staa3787}, \href
  {https://ui.adsabs.harvard.edu/abs/2021MNRAS.501.2934C} {501, 2934}

\bibitem[\protect\citeauthoryear{{D'Alessio} et~al.,}{{D'Alessio}
  et~al.}{2005}]{Dalessio2005}
{D'Alessio} P.,  et~al., 2005, \mn@doi [\apj] {10.1086/427490}, \href
  {http://adsabs.harvard.edu/abs/2005ApJ...621..461D} {621, 461}

\bibitem[\protect\citeauthoryear{{Davies} et~al.,}{{Davies}
  et~al.}{2020}]{Davies2020}
{Davies} C.~L.,  et~al., 2020, \mn@doi [\apj] {10.3847/1538-4357/ab93c1}, \href
  {https://ui.adsabs.harvard.edu/abs/2020ApJ...897...31D} {897, 31}

\bibitem[\protect\citeauthoryear{{Doi} \& {Kataoka}}{{Doi} \&
  {Kataoka}}{2021}]{Doi2021}
{Doi} K.,  {Kataoka} A.,  2021, \mn@doi [\apj] {10.3847/1538-4357/abe5a6},
  \href {https://ui.adsabs.harvard.edu/abs/2021ApJ...912..164D} {912, 164}

\bibitem[\protect\citeauthoryear{{Dong} et~al.,}{{Dong}
  et~al.}{2018}]{Dong2018_MWC758}
{Dong} R.,  et~al., 2018, \mn@doi [\apj] {10.3847/1538-4357/aac6cb}, \href
  {https://ui.adsabs.harvard.edu/abs/2018ApJ...860..124D} {860, 124}

\bibitem[\protect\citeauthoryear{{Dubrulle}, {Morfill}  \&
  {Sterzik}}{{Dubrulle} et~al.}{1995}]{Dubrulle1995}
{Dubrulle} B.,  {Morfill} G.,   {Sterzik} M.,  1995, \mn@doi [\icarus]
  {10.1006/icar.1995.1058}, \href
  {https://ui.adsabs.harvard.edu/abs/1995Icar..114..237D} {114, 237}

\bibitem[\protect\citeauthoryear{{Duch{\^e}ne} et~al.,}{{Duch{\^e}ne}
  et~al.}{2010}]{Duchene2010}
{Duch{\^e}ne} G.,  et~al., 2010, \mn@doi [\apj] {10.1088/0004-637X/712/1/112},
  \href {http://cdsads.u-strasbg.fr/abs/2010ApJ...712..112D} {712, 112}

\bibitem[\protect\citeauthoryear{{Duch{\^e}ne} et~al.,}{{Duch{\^e}ne}
  et~al.}{2024}]{Duchene2024}
{Duch{\^e}ne} G.,  et~al., 2024, \mn@doi [\aj] {10.3847/1538-3881/acf9a7},
  \href {https://ui.adsabs.harvard.edu/abs/2024AJ....167...77D} {167, 77}

\bibitem[\protect\citeauthoryear{{Dullemond} \& {Monnier}}{{Dullemond} \&
  {Monnier}}{2010}]{Dullemond2010}
{Dullemond} C.~P.,  {Monnier} J.~D.,  2010, \mn@doi [\araa]
  {10.1146/annurev-astro-081309-130932}, \href
  {https://ui.adsabs.harvard.edu/abs/2010ARA&A..48..205D} {48, 205}

\bibitem[\protect\citeauthoryear{{Dullemond}, {Dominik}  \&
  {Natta}}{{Dullemond} et~al.}{2001}]{Dullemond2001}
{Dullemond} C.~P.,  {Dominik} C.,   {Natta} A.,  2001, \mn@doi [\apj]
  {10.1086/323057}, \href {http://adsabs.harvard.edu/abs/2001ApJ...560..957D}
  {560, 957}

\bibitem[\protect\citeauthoryear{{Flaherty}, {Hughes}, {Rosenfeld}, {Andrews},
  {Chiang}, {Simon}, {Kerzner}  \& {Wilner}}{{Flaherty}
  et~al.}{2015}]{Flaherty2015}
{Flaherty} K.~M.,  {Hughes} A.~M.,  {Rosenfeld} K.~A.,  {Andrews} S.~M.,
  {Chiang} E.,  {Simon} J.~B.,  {Kerzner} S.,   {Wilner} D.~J.,  2015, \mn@doi
  [\apj] {10.1088/0004-637X/813/2/99}, \href
  {https://ui.adsabs.harvard.edu/abs/2015ApJ...813...99F} {813, 99}

\bibitem[\protect\citeauthoryear{{Flaherty} et~al.,}{{Flaherty}
  et~al.}{2017}]{Flaherty2017}
{Flaherty} K.~M.,  et~al., 2017, \mn@doi [\apj] {10.3847/1538-4357/aa79f9},
  \href {https://ui.adsabs.harvard.edu/abs/2017ApJ...843..150F} {843, 150}

\bibitem[\protect\citeauthoryear{{Flock}, {Fromang}, {Turner}  \&
  {Benisty}}{{Flock} et~al.}{2017}]{Flock2017}
{Flock} M.,  {Fromang} S.,  {Turner} N.~J.,   {Benisty} M.,  2017, \mn@doi
  [\apj] {10.3847/1538-4357/835/2/230}, \href
  {https://ui.adsabs.harvard.edu/abs/2017ApJ...835..230F} {835, 230}

\bibitem[\protect\citeauthoryear{{Foreman-Mackey}, {Hogg}, {Lang}  \&
  {Goodman}}{{Foreman-Mackey} et~al.}{2013}]{emcee}
{Foreman-Mackey} D.,  {Hogg} D.~W.,  {Lang} D.,   {Goodman} J.,  2013, \mn@doi
  [\pasp] {10.1086/670067}, \href
  {http://adsabs.harvard.edu/abs/2013PASP..125..306F} {125, 306}

\bibitem[\protect\citeauthoryear{{Francis} \& {van der Marel}}{{Francis} \&
  {van der Marel}}{2020}]{Francis2020}
{Francis} L.,  {van der Marel} N.,  2020, \mn@doi [\apj]
  {10.3847/1538-4357/ab7b63}, \href
  {https://ui.adsabs.harvard.edu/abs/2020ApJ...892..111F} {892, 111}

\bibitem[\protect\citeauthoryear{{Fromang} \& {Nelson}}{{Fromang} \&
  {Nelson}}{2009}]{Fromang2009}
{Fromang} S.,  {Nelson} R.~P.,  2009, \mn@doi [\aap]
  {10.1051/0004-6361/200811220}, \href
  {https://ui.adsabs.harvard.edu/abs/2009A&A...496..597F} {496, 597}

\bibitem[\protect\citeauthoryear{{Fukagawa} et~al.,}{{Fukagawa}
  et~al.}{2013}]{Fukagawa2013}
{Fukagawa} M.,  et~al., 2013, \mn@doi [\pasj] {10.1093/pasj/65.6.L14}, \href
  {https://ui.adsabs.harvard.edu/abs/2013PASJ...65L..14F} {65, L14}

\bibitem[\protect\citeauthoryear{{Gaia Collaboration}, {Vallenari, A.}, {Brown,
  A.G.A.}, {Prusti, T.}  \& {et al.}}{{Gaia Collaboration}
  et~al.}{2022}]{GaiaDR3}
{Gaia Collaboration} {Vallenari, A.} {Brown, A.G.A.} {Prusti, T.}  {et al.}
  2022, \mn@doi [A\&A] {10.1051/0004-6361/202243940}

\bibitem[\protect\citeauthoryear{{Garufi} et~al.,}{{Garufi}
  et~al.}{2019}]{Garufi2019}
{Garufi} A.,  et~al., 2019, \mn@doi [\aap] {10.1051/0004-6361/201935546}, \href
  {https://ui.adsabs.harvard.edu/abs/2019A&A...628A..68G} {628, A68}

\bibitem[\protect\citeauthoryear{{Garufi} et~al.,}{{Garufi}
  et~al.}{2022}]{Garufi2022}
{Garufi} A.,  et~al., 2022, \mn@doi [\aap] {10.1051/0004-6361/202141692}, \href
  {https://ui.adsabs.harvard.edu/abs/2022A&A...658A.137G} {658, A137}

\bibitem[\protect\citeauthoryear{{Ginski} et~al.,}{{Ginski}
  et~al.}{2016}]{Ginski2016}
{Ginski} C.,  et~al., 2016, \mn@doi [\aap] {10.1051/0004-6361/201629265}, \href
  {https://ui.adsabs.harvard.edu/abs/2016A&A...595A.112G} {595, A112}

\bibitem[\protect\citeauthoryear{{Godon} \& {Livio}}{{Godon} \&
  {Livio}}{2000}]{Godon2000}
{Godon} P.,  {Livio} M.,  2000, \mn@doi [\apj] {10.1086/309019}, \href
  {https://ui.adsabs.harvard.edu/abs/2000ApJ...537..396G} {537, 396}

\bibitem[\protect\citeauthoryear{{Guerra-Alvarado} et~al.,}{{Guerra-Alvarado}
  et~al.}{2024}]{GuerraAlvarado2024}
{Guerra-Alvarado} O.~M.,  et~al., 2024, \mn@doi [arXiv e-prints]
  {10.48550/arXiv.2404.04164}, \href
  {https://ui.adsabs.harvard.edu/abs/2024arXiv240404164G} {p. arXiv:2404.04164}

\bibitem[\protect\citeauthoryear{{Guidi} et~al.,}{{Guidi}
  et~al.}{2022}]{Guidi2022}
{Guidi} G.,  et~al., 2022, \mn@doi [\aap] {10.1051/0004-6361/202142303}, \href
  {https://ui.adsabs.harvard.edu/abs/2022A&A...664A.137G} {664, A137}

\bibitem[\protect\citeauthoryear{{Harsono} et~al.,}{{Harsono}
  et~al.}{2024}]{Harsono2024}
{Harsono} D.,  et~al., 2024, \mn@doi [\apj] {10.3847/1538-4357/ad0835}, \href
  {https://ui.adsabs.harvard.edu/abs/2024ApJ...961...28H} {961, 28}

\bibitem[\protect\citeauthoryear{{Hashimoto}, {Muto}, {Dong}, {Liu}, {van der
  Marel}, {Francis}, {Hasegawa}  \& {Tsukagoshi}}{{Hashimoto}
  et~al.}{2021}]{Hashimoto2021}
{Hashimoto} J.,  {Muto} T.,  {Dong} R.,  {Liu} H.~B.,  {van der Marel} N.,
  {Francis} L.,  {Hasegawa} Y.,   {Tsukagoshi} T.,  2021, \mn@doi [\apj]
  {10.3847/1538-4357/abe59f}, \href
  {https://ui.adsabs.harvard.edu/abs/2021ApJ...911....5H} {911, 5}

\bibitem[\protect\citeauthoryear{{Herczeg} \& {Hillenbrand}}{{Herczeg} \&
  {Hillenbrand}}{2014}]{Herczeg2014}
{Herczeg} G.~J.,  {Hillenbrand} L.~A.,  2014, \mn@doi [\apj]
  {10.1088/0004-637X/786/2/97}, \href
  {https://ui.adsabs.harvard.edu/abs/2014ApJ...786...97H} {786, 97}

\bibitem[\protect\citeauthoryear{{Huang} et~al.,}{{Huang}
  et~al.}{2018}]{Huang2018}
{Huang} J.,  et~al., 2018, \mn@doi [\apj] {10.3847/1538-4357/aaa1e7}, \href
  {https://ui.adsabs.harvard.edu/abs/2018ApJ...852..122H} {852, 122}

\bibitem[\protect\citeauthoryear{{Isella} \& {Natta}}{{Isella} \&
  {Natta}}{2005}]{Isella2005}
{Isella} A.,  {Natta} A.,  2005, \mn@doi [\aap] {10.1051/0004-6361:20052773},
  \href {http://adsabs.harvard.edu/abs/2005A%26A...438..899I} {438, 899}

\bibitem[\protect\citeauthoryear{{Isella}, {P{\'e}rez}, {Carpenter}, {Ricci},
  {Andrews}  \& {Rosenfeld}}{{Isella} et~al.}{2013}]{Isella2013}
{Isella} A.,  {P{\'e}rez} L.~M.,  {Carpenter} J.~M.,  {Ricci} L.,  {Andrews}
  S.,   {Rosenfeld} K.,  2013, \mn@doi [\apj] {10.1088/0004-637X/775/1/30},
  \href {https://ui.adsabs.harvard.edu/abs/2013ApJ...775...30I} {775, 30}

\bibitem[\protect\citeauthoryear{{Izquierdo}, {Testi}, {Facchini}, {Rosotti}
  \& {van Dishoeck}}{{Izquierdo} et~al.}{2021}]{Izquierdo2021}
{Izquierdo} A.~F.,  {Testi} L.,  {Facchini} S.,  {Rosotti} G.~P.,   {van
  Dishoeck} E.~F.,  2021, \mn@doi [\aap] {10.1051/0004-6361/202140779}, \href
  {https://ui.adsabs.harvard.edu/abs/2021A&A...650A.179I} {650, A179}

\bibitem[\protect\citeauthoryear{{Jennings}, {Booth}, {Tazzari}, {Rosotti}  \&
  {Clarke}}{{Jennings} et~al.}{2020}]{frank}
{Jennings} J.,  {Booth} R.~A.,  {Tazzari} M.,  {Rosotti} G.~P.,   {Clarke}
  C.~J.,  2020, \mn@doi [\mnras] {10.1093/mnras/staa1365}, \href
  {https://ui.adsabs.harvard.edu/abs/2020MNRAS.495.3209J} {495, 3209}

\bibitem[\protect\citeauthoryear{{Jennings}, {Tazzari}, {Clarke}, {Booth}  \&
  {Rosotti}}{{Jennings} et~al.}{2022}]{Jennings2022}
{Jennings} J.,  {Tazzari} M.,  {Clarke} C.~J.,  {Booth} R.~A.,   {Rosotti}
  G.~P.,  2022, \mn@doi [\mnras] {10.1093/mnras/stac1770}, \href
  {https://ui.adsabs.harvard.edu/abs/2022MNRAS.514.6053J} {514, 6053}

\bibitem[\protect\citeauthoryear{{Johansen} \& {Lambrechts}}{{Johansen} \&
  {Lambrechts}}{2017}]{Johansen2017}
{Johansen} A.,  {Lambrechts} M.,  2017, \mn@doi [Annual Review of Earth and
  Planetary Sciences] {10.1146/annurev-earth-063016-020226}, \href
  {https://ui.adsabs.harvard.edu/abs/2017AREPS..45..359J} {45, 359}

\bibitem[\protect\citeauthoryear{{Kama}, {Min}  \& {Dominik}}{{Kama}
  et~al.}{2009}]{Kama2009}
{Kama} M.,  {Min} M.,   {Dominik} C.,  2009, \mn@doi [\aap]
  {10.1051/0004-6361/200912068}, \href
  {http://adsabs.harvard.edu/abs/2009A%26A...506.1199K} {506, 1199}

\bibitem[\protect\citeauthoryear{{Kluska} et~al.,}{{Kluska}
  et~al.}{2020}]{Kluska2020}
{Kluska} J.,  et~al., 2020, \mn@doi [\aap] {10.1051/0004-6361/201833774}, \href
  {https://ui.adsabs.harvard.edu/abs/2020A&A...636A.116K} {636, A116}

\bibitem[\protect\citeauthoryear{{Langlois} et~al.,}{{Langlois}
  et~al.}{2018}]{Langlois2018}
{Langlois} M.,  et~al., 2018, \mn@doi [\aap] {10.1051/0004-6361/201731624},
  \href {https://ui.adsabs.harvard.edu/abs/2018A&A...614A..88L} {614, A88}

\bibitem[\protect\citeauthoryear{Law et~al.,}{Law et~al.}{2022}]{Law2022}
Law C.~J.,  et~al., 2022, \mn@doi [The Astrophysical Journal]
  {10.3847/1538-4357/ac6c02}, 932, 114

\bibitem[\protect\citeauthoryear{{Law} et~al.,}{{Law} et~al.}{2023}]{Law2023}
{Law} C.~J.,  et~al., 2023, \mn@doi [\apj] {10.3847/1538-4357/acb3c4}, \href
  {https://ui.adsabs.harvard.edu/abs/2023ApJ...948...60L} {948, 60}

\bibitem[\protect\citeauthoryear{{Lazareff} et~al.,}{{Lazareff}
  et~al.}{2017}]{Lazareff2017}
{Lazareff} B.,  et~al., 2017, \mn@doi [\aap] {10.1051/0004-6361/201629305},
  \href {https://ui.adsabs.harvard.edu/abs/2017A&A...599A..85L} {599, A85}

\bibitem[\protect\citeauthoryear{{Long} et~al.,}{{Long}
  et~al.}{2018}]{Long2018}
{Long} F.,  et~al., 2018, \mn@doi [\apj] {10.3847/1538-4357/aae8e1}, \href
  {https://ui.adsabs.harvard.edu/abs/2018ApJ...869...17L} {869, 17}

\bibitem[\protect\citeauthoryear{{Mac{\'\i}as}, {Guerra-Alvarado},
  {Carrasco-Gonz{\'a}lez}, {Ribas}, {Espaillat}, {Huang}  \&
  {Andrews}}{{Mac{\'\i}as} et~al.}{2021}]{Macias2021}
{Mac{\'\i}as} E.,  {Guerra-Alvarado} O.,  {Carrasco-Gonz{\'a}lez} C.,  {Ribas}
  {\'A}.,  {Espaillat} C.~C.,  {Huang} J.,   {Andrews} S.~M.,  2021, \mn@doi
  [\aap] {10.1051/0004-6361/202039812}, \href
  {https://ui.adsabs.harvard.edu/abs/2021A&A...648A..33M} {648, A33}

\bibitem[\protect\citeauthoryear{{Meheut}, {Meliani}, {Varniere}  \&
  {Benz}}{{Meheut} et~al.}{2012}]{Meheut2012}
{Meheut} H.,  {Meliani} Z.,  {Varniere} P.,   {Benz} W.,  2012, \mn@doi [\aap]
  {10.1051/0004-6361/201219794}, \href
  {https://ui.adsabs.harvard.edu/abs/2012A&A...545A.134M} {545, A134}

\bibitem[\protect\citeauthoryear{{M{\'e}nard} et~al.,}{{M{\'e}nard}
  et~al.}{2020}]{Menard2020}
{M{\'e}nard} F.,  et~al., 2020, \mn@doi [\aap] {10.1051/0004-6361/202038356},
  \href {https://ui.adsabs.harvard.edu/abs/2020A&A...639L...1M} {639, L1}

\bibitem[\protect\citeauthoryear{{Min}, {Rab}, {Woitke}, {Dominik}  \&
  {M{\'e}nard}}{{Min} et~al.}{2016}]{Min2016}
{Min} M.,  {Rab} C.,  {Woitke} P.,  {Dominik} C.,   {M{\'e}nard} F.,  2016,
  \mn@doi [\aap] {10.1051/0004-6361/201526048}, \href
  {https://ui.adsabs.harvard.edu/abs/2016A&A...585A..13M} {585, A13}

\bibitem[\protect\citeauthoryear{{Natta}, {Prusti}, {Neri}, {Wooden}, {Grinin}
  \& {Mannings}}{{Natta} et~al.}{2001}]{Natta2001}
{Natta} A.,  {Prusti} T.,  {Neri} R.,  {Wooden} D.,  {Grinin} V.~P.,
  {Mannings} V.,  2001, \mn@doi [\aap] {10.1051/0004-6361:20010334}, \href
  {https://ui.adsabs.harvard.edu/abs/2001A&A...371..186N} {371, 186}

\bibitem[\protect\citeauthoryear{{Orihara} et~al.,}{{Orihara}
  et~al.}{2023}]{Orihara2023}
{Orihara} R.,  et~al., 2023, \mn@doi [\pasj] {10.1093/pasj/psad009}, \href
  {https://ui.adsabs.harvard.edu/abs/2023PASJ...75..424O} {75, 424}

\bibitem[\protect\citeauthoryear{{Pinilla}, {Birnstiel}, {Ricci}, {Dullemond},
  {Uribe}, {Testi}  \& {Natta}}{{Pinilla} et~al.}{2012}]{Pinilla2012}
{Pinilla} P.,  {Birnstiel} T.,  {Ricci} L.,  {Dullemond} C.~P.,  {Uribe} A.~L.,
   {Testi} L.,   {Natta} A.,  2012, \mn@doi [\aap]
  {10.1051/0004-6361/201118204}, \href
  {http://adsabs.harvard.edu/abs/2012A%26A...538A.114P} {538, A114}

\bibitem[\protect\citeauthoryear{{Pinilla} et~al.,}{{Pinilla}
  et~al.}{2022}]{Pinilla2022}
{Pinilla} P.,  et~al., 2022, \mn@doi [\aap] {10.1051/0004-6361/202243704},
  \href {https://ui.adsabs.harvard.edu/abs/2022A&A...665A.128P} {665, A128}

\bibitem[\protect\citeauthoryear{{Pinte}, {M{\'e}nard}, {Duch{\^e}ne}  \&
  {Bastien}}{{Pinte} et~al.}{2006}]{MCFOST}
{Pinte} C.,  {M{\'e}nard} F.,  {Duch{\^e}ne} G.,   {Bastien} P.,  2006, \mn@doi
  [\aap] {10.1051/0004-6361:20053275}, \href
  {http://adsabs.harvard.edu/abs/2006A%26A...459..797P} {459, 797}

\bibitem[\protect\citeauthoryear{{Pinte}, {Harries}, {Min}, {Watson},
  {Dullemond}, {Woitke}, {M{\'e}nard}  \& {Dur{\'a}n-Rojas}}{{Pinte}
  et~al.}{2009}]{MCFOST2}
{Pinte} C.,  {Harries} T.~J.,  {Min} M.,  {Watson} A.~M.,  {Dullemond} C.~P.,
  {Woitke} P.,  {M{\'e}nard} F.,   {Dur{\'a}n-Rojas} M.~C.,  2009, \mn@doi
  [\aap] {10.1051/0004-6361/200811555}, \href
  {http://adsabs.harvard.edu/abs/2009A%26A...498..967P} {498, 967}

\bibitem[\protect\citeauthoryear{{Pinte}, {Dent}, {M{\'e}nard}, {Hales},
  {Hill}, {Cortes}  \& {de Gregorio-Monsalvo}}{{Pinte}
  et~al.}{2016}]{Pinte2016}
{Pinte} C.,  {Dent} W.~R.~F.,  {M{\'e}nard} F.,  {Hales} A.,  {Hill} T.,
  {Cortes} P.,   {de Gregorio-Monsalvo} I.,  2016, \mn@doi [\apj]
  {10.3847/0004-637X/816/1/25}, \href
  {https://ui.adsabs.harvard.edu/abs/2016ApJ...816...25P} {816, 25}

\bibitem[\protect\citeauthoryear{{Pinte} et~al.,}{{Pinte}
  et~al.}{2018}]{Pinte2018}
{Pinte} C.,  et~al., 2018, \mn@doi [\apjl] {10.3847/2041-8213/aac6dc}, \href
  {https://ui.adsabs.harvard.edu/abs/2018ApJ...860L..13P} {860, L13}

\bibitem[\protect\citeauthoryear{{Pizzati}, {Rosotti}  \& {Tabone}}{{Pizzati}
  et~al.}{2023}]{Pizzati2023}
{Pizzati} E.,  {Rosotti} G.~P.,   {Tabone} B.,  2023, \mn@doi [\mnras]
  {10.1093/mnras/stad2057}, \href
  {https://ui.adsabs.harvard.edu/abs/2023MNRAS.524.3184P} {524, 3184}

\bibitem[\protect\citeauthoryear{{Ragusa}, {Dipierro}, {Lodato}, {Laibe}  \&
  {Price}}{{Ragusa} et~al.}{2017}]{Ragusa2017}
{Ragusa} E.,  {Dipierro} G.,  {Lodato} G.,  {Laibe} G.,   {Price} D.~J.,  2017,
  \mn@doi [\mnras] {10.1093/mnras/stw2456}, \href
  {https://ui.adsabs.harvard.edu/abs/2017MNRAS.464.1449R} {464, 1449}

\bibitem[\protect\citeauthoryear{{Ragusa}, {Alexander}, {Calcino}, {Hirsh}  \&
  {Price}}{{Ragusa} et~al.}{2020}]{Ragusa2020}
{Ragusa} E.,  {Alexander} R.,  {Calcino} J.,  {Hirsh} K.,   {Price} D.~J.,
  2020, \mn@doi [\mnras] {10.1093/mnras/staa2954}, \href
  {https://ui.adsabs.harvard.edu/abs/2020MNRAS.499.3362R} {499, 3362}

\bibitem[\protect\citeauthoryear{{Rau} \& {Cornwell}}{{Rau} \&
  {Cornwell}}{2011}]{Rau2011}
{Rau} U.,  {Cornwell} T.~J.,  2011, \mn@doi [\aap]
  {10.1051/0004-6361/201117104}, \href
  {https://ui.adsabs.harvard.edu/abs/2011A&A...532A..71R} {532, A71}

\bibitem[\protect\citeauthoryear{{Ribas}, {Espaillat}, {Mac{\'\i}as}  \&
  {Sarro}}{{Ribas} et~al.}{2020}]{Ribas2020}
{Ribas} {\'A}.,  {Espaillat} C.~C.,  {Mac{\'\i}as} E.,   {Sarro} L.~M.,  2020,
  \mn@doi [\aap] {10.1051/0004-6361/202038352}, \href
  {https://ui.adsabs.harvard.edu/abs/2020A&A...642A.171R} {642, A171}

\bibitem[\protect\citeauthoryear{{Ribas} et~al.,}{{Ribas}
  et~al.}{2023}]{Ribas2023}
{Ribas} {\'A}.,  et~al., 2023, \mn@doi [\aap] {10.1051/0004-6361/202245637},
  \href {https://ui.adsabs.harvard.edu/abs/2023A&A...673A..77R} {673, A77}

\bibitem[\protect\citeauthoryear{{Rich} et~al.,}{{Rich}
  et~al.}{2019}]{Rich2019}
{Rich} E.~A.,  et~al., 2019, \mn@doi [\apj] {10.3847/1538-4357/ab0f3b}, \href
  {https://ui.adsabs.harvard.edu/abs/2019ApJ...875...38R} {875, 38}

\bibitem[\protect\citeauthoryear{{Rilinger}, {Espaillat}, {Xin}, {Ribas},
  {Mac{\'\i}as}  \& {Luettgen}}{{Rilinger} et~al.}{2023}]{Rilinger2023}
{Rilinger} A.~M.,  {Espaillat} C.~C.,  {Xin} Z.,  {Ribas} {\'A}.,
  {Mac{\'\i}as} E.,   {Luettgen} S.,  2023, \mn@doi [\apj]
  {10.3847/1538-4357/aca905}, \href
  {https://ui.adsabs.harvard.edu/abs/2023ApJ...944...66R} {944, 66}

\bibitem[\protect\citeauthoryear{{Shakura} \& {Sunyaev}}{{Shakura} \&
  {Sunyaev}}{1973}]{Shakura1973}
{Shakura} N.~I.,  {Sunyaev} R.~A.,  1973, \aap, \href
  {http://adsabs.harvard.edu/abs/1973A%26A....24..337S} {24, 337}

\bibitem[\protect\citeauthoryear{{Sierra} et~al.,}{{Sierra}
  et~al.}{2021}]{Sierra2021}
{Sierra} A.,  et~al., 2021, \mn@doi [\apjs] {10.3847/1538-4365/ac1431}, \href
  {https://ui.adsabs.harvard.edu/abs/2021ApJS..257...14S} {257, 14}

\bibitem[\protect\citeauthoryear{{Soon}, {Hanawa}, {Muto}, {Tsukagoshi}  \&
  {Momose}}{{Soon} et~al.}{2017}]{Soon2017}
{Soon} K.-L.,  {Hanawa} T.,  {Muto} T.,  {Tsukagoshi} T.,   {Momose} M.,  2017,
  \mn@doi [\pasj] {10.1093/pasj/psx007}, \href
  {https://ui.adsabs.harvard.edu/abs/2017PASJ...69...34S} {69, 34}

\bibitem[\protect\citeauthoryear{{Tang}, {Guilloteau}, {Pi{\'e}tu}, {Dutrey},
  {Ohashi}  \& {Ho}}{{Tang} et~al.}{2012}]{Tang2012}
{Tang} Y.~W.,  {Guilloteau} S.,  {Pi{\'e}tu} V.,  {Dutrey} A.,  {Ohashi} N.,
  {Ho} P.~T.~P.,  2012, \mn@doi [\aap] {10.1051/0004-6361/201219414}, \href
  {https://ui.adsabs.harvard.edu/abs/2012A&A...547A..84T} {547, A84}

\bibitem[\protect\citeauthoryear{{Tazzari}, {Beaujean}  \& {Testi}}{{Tazzari}
  et~al.}{2018}]{galario}
{Tazzari} M.,  {Beaujean} F.,   {Testi} L.,  2018, \mn@doi [\mnras]
  {10.1093/mnras/sty409}, \href
  {https://ui.adsabs.harvard.edu/abs/2018MNRAS.476.4527T} {476, 4527}

\bibitem[\protect\citeauthoryear{{Teague}, {Bae}, {Huang}  \&
  {Bergin}}{{Teague} et~al.}{2019}]{Teague2019}
{Teague} R.,  {Bae} J.,  {Huang} J.,   {Bergin} E.~A.,  2019, \mn@doi [\apjl]
  {10.3847/2041-8213/ab4a83}, \href
  {https://ui.adsabs.harvard.edu/abs/2019ApJ...884L..56T} {884, L56}

\bibitem[\protect\citeauthoryear{{Tripathi}, {Andrews}, {Birnstiel}  \&
  {Wilner}}{{Tripathi} et~al.}{2017}]{Tripathi2017}
{Tripathi} A.,  {Andrews} S.~M.,  {Birnstiel} T.,   {Wilner} D.~J.,  2017,
  \mn@doi [\apj] {10.3847/1538-4357/aa7c62}, \href
  {https://ui.adsabs.harvard.edu/abs/2017ApJ...845...44T} {845, 44}

\bibitem[\protect\citeauthoryear{{Valeg{\r{a}}rd} et~al.,}{{Valeg{\r{a}}rd}
  et~al.}{2022}]{Valegard2022}
{Valeg{\r{a}}rd} P.~G.,  et~al., 2022, \mn@doi [\aap]
  {10.1051/0004-6361/202244001}, \href
  {https://ui.adsabs.harvard.edu/abs/2022A&A...668A..25V} {668, A25}

\bibitem[\protect\citeauthoryear{{Villenave} et~al.,}{{Villenave}
  et~al.}{2020}]{Villenave2020}
{Villenave} M.,  et~al., 2020, \mn@doi [\aap] {10.1051/0004-6361/202038087},
  \href {https://ui.adsabs.harvard.edu/abs/2020A&A...642A.164V} {642, A164}

\bibitem[\protect\citeauthoryear{{Villenave} et~al.,}{{Villenave}
  et~al.}{2022}]{Villenave2022}
{Villenave} M.,  et~al., 2022, \mn@doi [\apj] {10.3847/1538-4357/ac5fae}, \href
  {https://ui.adsabs.harvard.edu/abs/2022ApJ...930...11V} {930, 11}

\bibitem[\protect\citeauthoryear{{Villenave} et~al.,}{{Villenave}
  et~al.}{2024}]{Villenave2024}
{Villenave} M.,  et~al., 2024, \mn@doi [\apj] {10.3847/1538-4357/ad0c4b}, \href
  {https://ui.adsabs.harvard.edu/abs/2024ApJ...961...95V} {961, 95}

\bibitem[\protect\citeauthoryear{{Weber} et~al.,}{{Weber}
  et~al.}{2023}]{Weber2023}
{Weber} P.,  et~al., 2023, \mn@doi [\mnras] {10.1093/mnras/stac3478}, \href
  {https://ui.adsabs.harvard.edu/abs/2023MNRAS.518.5620W} {518, 5620}

\bibitem[\protect\citeauthoryear{{Woitke} et~al.,}{{Woitke}
  et~al.}{2016}]{Woitke2016}
{Woitke} P.,  et~al., 2016, \mn@doi [\aap] {10.1051/0004-6361/201526538}, \href
  {http://adsabs.harvard.edu/abs/2016A%26A...586A.103W} {586, A103}

\bibitem[\protect\citeauthoryear{{Wolff} et~al.,}{{Wolff}
  et~al.}{2017}]{Wolff2017}
{Wolff} S.~G.,  et~al., 2017, \mn@doi [\apj] {10.3847/1538-4357/aa9981}, \href
  {https://ui.adsabs.harvard.edu/abs/2017ApJ...851...56W} {851, 56}

\bibitem[\protect\citeauthoryear{{Xin}, {Espaillat}, {Rilinger}, {Ribas}  \&
  {Mac{\'\i}as}}{{Xin} et~al.}{2023}]{Xin2023}
{Xin} Z.,  {Espaillat} C.~C.,  {Rilinger} A.~M.,  {Ribas} {\'A}.,
  {Mac{\'\i}as} E.,  2023, \mn@doi [\apj] {10.3847/1538-4357/aca52b}, \href
  {https://ui.adsabs.harvard.edu/abs/2023ApJ...942....4X} {942, 4}

\bibitem[\protect\citeauthoryear{{Youdin} \& {Goodman}}{{Youdin} \&
  {Goodman}}{2005}]{Youdin2005}
{Youdin} A.~N.,  {Goodman} J.,  2005, \mn@doi [\apj] {10.1086/426895}, \href
  {https://ui.adsabs.harvard.edu/abs/2005ApJ...620..459Y} {620, 459}

\bibitem[\protect\citeauthoryear{{van Terwisga} et~al.,}{{van Terwisga}
  et~al.}{2020}]{vanTerwisga2020}
{van Terwisga} S.~E.,  et~al., 2020, \mn@doi [\aap]
  {10.1051/0004-6361/201937403}, \href
  {https://ui.adsabs.harvard.edu/abs/2020A&A...640A..27V} {640, A27}

\bibitem[\protect\citeauthoryear{{van der Marel} et~al.,}{{van der Marel}
  et~al.}{2013}]{vanderMarel2013}
{van der Marel} N.,  et~al., 2013, \mn@doi [Science] {10.1126/science.1236770},
  \href {http://adsabs.harvard.edu/abs/2013Sci...340.1199V} {340, 1199}

\bibitem[\protect\citeauthoryear{{van der Plas} et~al.,}{{van der Plas}
  et~al.}{2017}]{vanderPlas2017}
{van der Plas} G.,  et~al., 2017, \mn@doi [\aap] {10.1051/0004-6361/201629523},
  \href {http://adsabs.harvard.edu/abs/2017A%26A...597A..32V} {597, A32}

\makeatother
\end{thebibliography}

% Don't change these lines
\bsp	% typesetting comment
\label{lastpage}
\end{document}